\documentclass[10pt]{article}
\usepackage{amsmath, amssymb}
\usepackage[mathscr]{eucal}
\usepackage[usenames,dvipsnames]{color}

\usepackage{graphicx}          

\allowdisplaybreaks

\newfont{\bssfont}{cmssbx10}

\def\ii{{\rm i\,}}

\def\fracpd#1#2{\frac{\partial #1}{\partial #2}} 
\def\fracpartial#1{\frac{\partial}{\partial #1}} 

\def\unomasuno{{1\!+\!1}}
\def\smallonehalf{\frac{{}_1}{{}^2}}

\def\={\!=\!}
\def\>{\!>\!}
\def\<{\!<\!}

\def\k{\kappa}
\def\ki{\kappa_1}
\def\kii{\kappa_2}

\def\quadrant{{\vert\kern-.15em\underline{\, \cdot\, } \,} }

\def\CKspace{{\textbf{\itshape S}}^{\,2}_{\ki\![\kii]}}

\def\CKC{\,\mathop{\kern-.15em\rm C}\nolimits}    
\def\CKS{\,\mathop{\kern-.15em\rm S}\nolimits}    
\def\CKT{\,\mathop{\kern-.15em\rm T}\nolimits}    
\def\CKV{\,\mathop{\kern-.15em\rm V}\nolimits}    
\def\CKtoV{\,\mathop{\kern-.15em\rm W}\nolimits}    

\def\CKArcT{\,\mathop{\kern-.15em\rm ArcT}\nolimits}    

\def\CKc{\CKC_{\ki}\!}
\def\CKs{\CKS_{\ki}\!}
\def\CKt{\CKT_{\ki}\!}

\def\CKcc{\CKC_{\kii}\!}
\def\CKss{\CKS_{\kii}\!}

\def\CKccc{\CKC_{\ki\!\kii}\!}
\def\CKsss{\CKS_{\ki\!\kii}\!}
\def\CKttt{\CKT_{\ki\!\kii}\!}

\def\CKcSq{\CKC^2_{\ki}\!}  
\def\CKsSq{\CKS^2_{\ki}\!}  
\def\CKtSq{\CKT^2_{\ki}\!}

\def\CKccSq{\CKC^2_{\kii}\!}
\def\CKssSq{\CKS^2_{\kii}\!}

\def\CKcccSq{\CKC^2_{\ki\!\kii}\!}

\def\CKtttSq{\CKT^2_{\ki\!\kii}\!}

\def\CKCr{\CKc(r)}    
\def\CKCu{\CKc(u)}  \def\CKCuSq{\CKcSq(u)}
  \def\CKCxSq{\CKcSq(x)}
\def\CKSr{\CKs(r)}  \def\CKSrSq{\CKsSq(r)}  
\def\CKSu{\CKs(u)}

\def\CKTr{\CKt(r)}  \def\CKTrSq{\CKtSq(r)}
\def\CKTu{\CKt(u)}  \def\CKTuSq{\CKtSq(u)}

\def\CKCphi{\CKcc(\phi)}  \def\CKCphiSq{\CKccSq(\phi)}
\def\CKSphi{\CKss(\phi)}  \def\CKSphiSq{\CKssSq(\phi)}

\def\CKCy{\CKccc(y)} \def\CKCySq{\CKcccSq(y)}
 
\def\CKSy{\CKsss(y)} 

\def\CKTy{\CKttt(y)} \def\CKTySq{\CKtttSq(y)}

\def\CKTa{\CKt(a)} \def\CKTaSq{\CKtSq(a)}
\def\CKTaTl{\CKt(\tilde{a})} \def\CKTaTlSq{\CKtSq(\tilde{a})}
 
\def\CKTb{\CKttt(b)} \def\CKTbSq{\CKtttSq(b)}

\def\CKP{\mathcal{P}}
\def\CKJ{\mathcal{J}}
\def\CKW{\mathcal{W}}

\def\CKF{\mathcal{F}}

\def\CKLag{\mathcal{L}}
\def\CKPot{\mathcal{V}}

\def\CKPoteff{\CKPot^{\rm eff}}

\def\f{\phi}

\def\uRegPar{\upsilon}

\def\IL{\relax{\rm I\kern-.18 em L}}

\input epsf
\begin{document}

\baselineskip14pt  

\pagestyle{myheadings}\markboth{Harmonic Oscillator on Riemannian 
\dots Cari\~nena et al.  }{Harmonic Oscillator on Riemannian \dots 
Cari\~nena et al.  }

\title{The harmonic oscillator on Riemannian and Lorentzian configuration
spaces of constant curvature 
}

\author{Jos\'e F. Cari\~nena$\dagger\,^{a)}$,
         Manuel F. Ra\~nada$\dagger\,^{b)}$  and
         Mariano Santander$\ddagger\,^{c)}$ \\[4pt]
$\dagger${\enskip}
    {\sl Departamento de F\'{\i}sica Te\'orica, Facultad de Ciencias}\\
    {\sl Universidad de Zaragoza, 50009 Zaragoza, Spain}  \\[2pt]
$\ddagger${\enskip}
    {\sl Departamento de F\'{\i}sica Te\'orica, Facultad de Ciencias}\\
    {\sl Universidad de Valladolid,  47011 Valladolid, Spain} }
\maketitle
\date{}

\begin{abstract}
The harmonic oscillator as a distinguished dynamical system can be
defined not only on the Euclidean plane but also on the sphere and
on the hyperbolic plane, and more generally on any configuration
space with constant curvature and with a metric of any signature,
either Riemannian (definite positive) or Lorentzian (indefinite).
In this paper we study the main properties of these `curved'
harmonic oscillators simultaneously on any such configuration
space, using a Cayley-Klein (CK) type approach, with two free
parameters $\ki, \kii$ which altogether correspond to the possible
values for curvature and signature type: the generic Riemannian
and Lorentzian spaces of constant curvature (sphere ${\bf S}^2$,
hyperbolic plane ${\bf H}^2$, AntiDeSitter sphere ${\bf
AdS}^{\unomasuno}$ and DeSitter sphere ${\bf dS}^{\unomasuno}$)
appear in this family, with the Euclidean and Minkowski spaces as
flat limits.

We solve the equations of motion for the `curved' harmonic
oscillator and obtain explicit expressions for the orbits by using
three different methods: first by direct integration, second by
obtaining the general CK version of the Binet's equation and
third, as a consequence of its superintegrable character. The
orbits are conics with centre at the potential origin in any CK
space, thereby extending this well known Euclidean property to any
constant curvature configuration space. The final part of the
article, that has a more geometric character, presents those
results of the theory of conics on spaces of constant curvature
which are pertinent.
\end{abstract}

\begin{quote}

The harmonic oscillator.  Integrability and Superintegrability.
Spaces of constant curvature. Central potentials. Conics on the
sphere and hyperbolic plane. Conics in Lorentzian spaces.

{\sl Running title:}{\enskip}
The harmonic oscillator on spaces with constant curvature.


\end{quote}
momentum maps, reduction
structure of phase space, integration methods
{\vfill}

\footnoterule
{\noindent\small
$^{a)}${\sl E-mail address:} {jfc@unizar.es} \\
$^{b)}${\sl E-mail address:} {mfran@unizar.es} \\
$^{c)}${\sl E-mail address:} {msn@fta.uva.es} }
\newpage




\section{Introduction}

In a sense this article can be considered as a sequel or
continuation of a previous paper \cite{CRS05} which was devoted to
the study of mechanical systems on Riemannian configuration spaces
with constant curvature $\kappa\ne 0$. Geodesic motion, the theory
of symmetries and general results on central potentials were
discussed in the first part of \cite{CRS05}, while in the second
part attention was focused on the Kepler problem in ${\bf S}^2$
and ${\bf H}^2$. Now, we present a similar analysis for the
harmonic oscillator, yet extending the scope so as to include also
the much less explored cases where the configuration space is a
{\it Lorentzian} manifold with constant curvature. We follow the
approach of \cite{CRS05}, which contains the fundamental ideas and
motivations, and also use the notation, ideas and results
discussed in \cite{RaS02b,RaSa03}.

The study of mechanical systems on Riemannian spaces has been
mainly done in connection with relativity and gravitation.
Nevertheless, before relativity, the study of both Kepler and
harmonic oscillator potentials in spaces of {\it constant
curvature} had also been done from the viewpoint of classical
nonrelativistic mechanics (see \cite{DoZi91} for an historical
account of the research made until the first years of the XX
century and references in \cite{CRS05} for more recent papers
including also some quantum problems on spaces with curvature as,
e.g., the hydrogen atom in a spherical or hyperbolic geometry). It
is interesting to point out that \cite{Lieb}, a book on geometry,
includes however a final chapter devoted to mechanics (the title
of this chapter was {\it ``Nichteuklidische Mechanik"} in the
original edition but was changed to {\it ``Mechanik und spezielle
Relativit\"atstheorie"} in the revised 1923 edition); in addition
to rather general properties linking geometry with mechanics, this
chapter contains the basics of a study of the harmonic  oscillator
in constant curvature; polar coordinates are used and the approach
is basically Newtonian.

It is well known that the Kepler problem and the harmonic
oscillator are the two more important superintegrable systems in
Euclidean space (see for instance the recent book \cite{Co03}),
and, as it was to be expected, they have `curved versions' which
remain superintegrable in spherical or hyperbolic configuration
spaces \cite{Sl00, Wozmis03}. This known property implicitly
underlies some classical papers as \cite{Sch40} or \cite{Hi79}. On
the other hand, much work has been recently done in the study of
superintegrable systems in non-Euclidean spaces \cite{BaW85,
BaIJ87, BaIJ90, BoDaK93, BoDK94, Gr90, KaMiPo97,KaKP01, KaKrMi05I,
KaKrMi05II, RaSa99, RS04, Sl00, TeTW01} and this fact has
intensified the interest for the study of the `curved' versions of
these two systems and their relations \cite{KaMiPo00, KaMP02,
RS02rmp33, BanachCP59}. A further step, which we take in this
paper, is to extend these studies also to the case where the
configuration space is itself a constant curvature Lorentzian
manifold. This case was definitely not taken into account in the
previous papers, and opens some views into a relatively unknown
field. For some work related to  dynamics in Lorentzian manifolds,
see \cite{Be06,  McLenSmiCoorMink, McLenSmi04, Szyd96}.

The three {\it classical} spaces with constant curvature $\k$, to
wit, the sphere ${\bf S}^2_\k$ with $\k>0$, Euclidean plane ${\bf
E}^2$ for $\k=0$ and hyperbolic plane ${\bf H}^2_\k$ for $\k<0$,
can be considered as the three different instances in the family
of homogeneous Riemannian manifolds $V_{\k}^2=({\bf S}^2_\k,\ {\bf
E}^2,\ {\bf H}^2_\k)$. A technique for  considering these three
spaces at the same time in a unique family, with the curvature
$\k$ as a parameter $\k\in\mathbb{R}$ was first introduced by
Weierstrass and Killing \cite{DoZi91} and lies at the origin of
the so-called Weierstrass model for the hyperbolic plane; for some
reason modern presentations usually restrict to the standard value
$\k=-1$, thus losing from direct view how some properties depend
on the curvature. If $\k$ is left explicitly, this allows
consideration of the `curved' harmonic oscillator (or Kepler
problem) on constant curvature spaces as arising from Lagrangians
depending on $\k$ as a parameter and understood as defined in a
generic space $V_{\k}^2$ (either $({\bf S}^2_\k,\ {\bf E}^2,\ {\bf
H}^2_\k)$) by using a unique $\k$-dependent expression. The
`curved' systems appear as a `$\k$-deformation' of the well known
Euclidean system and they can be defined without ambiguities
because superintegrability picks up essentially a unique
`$\k$-deformation' among the many (non-superintegrable) potentials
having the (Euclidean) Kepler or oscillator as its `$\k\to0$
limit'.

A convenient tool for the use of $\k$ as a parameter are the
following $\kappa$-trigonometric `Sine' and `Cosine' functions
\begin{equation}
\CKC_{\k}(x) :=\left\{
\begin{array}{l}
   \cos {\sqrt{\k}\, x} \cr
   1  \cr
\cosh {\sqrt{-\k}\, x}
\end{array}\right. ,
\qquad
\CKS_{\k}(x) :=\left\{
\begin{array}{ll}
     \frac{1}{\sqrt{\k}} \sin {\sqrt{\k}\, x} &\qquad  \k >0 \cr
   x &\qquad  \k  =0 \cr
\frac{1}{\sqrt{-\k}} \sinh {\sqrt{-\k}\, x} &\qquad   \k <0
\end{array}\right.,
\label{ckho07:CKDefSinCos}
\end{equation}
as well as the `Tangent' $\CKT_{\k}(x) =
\CKS_{\k}(x)/\CKC_{\k}(x)$. These functions allow us to write
$\k$-dependent expressions in a unified way for the whole previous
family of spaces $V_{\k}^2$ so that the computations, statements
and results are unified too.

This approach, used in some previous papers \cite{BHSS03, CRS05,
RaSa99, RaS02b, RaSa03}, has a potentiality beyond the unification
for the three constant curvature configuration spaces $V_{\k}^2$:
the mathematically natural frame for the `$\k$ as a parameter'
idea \cite{CRSS05, HBSS04, HeBa06,  SaSa05} involves not just one
single parameter $\k$, but {\it two parameters}, $\ki$ and $\kii$,
which correspond to a space $\CKspace$ with constant curvature
$\ki$ and whose metric is definite (Riemannian), degenerate or
indefinite (Lorentzian) according to the $\kii$ sign (see e.g.
\cite{BHOS93, CRS07a, ConformalHS02, ConformalCompactHS02}).

In more detail, the plan of this article is as follows: In Sec.\
II we analyse, in a single run,  the free geodesic motion on the
three Riemannian spaces $V^2_\k\equiv({\bf S}^2_\k,{\bf E}^2,{\bf
H}^2_\k)$,  and in the three Lorentzian spaces $L^2_\k\equiv({\bf
AdS}^{\unomasuno}_\k, {\bf M}^{\unomasuno}, {\bf
dS}^{\unomasuno}_\k)$ as well as the Killing vector fields in the
general $\CKspace$ and its correspondent Noether symmetries. We
have divided this section in two subsections, using respectively
geodesic polar coordinates and geodesic parallel coordinates.

Sec.\ III, that can be considered as the central part of the
article, is devoted to the study of the harmonic oscillator on the
general space $\CKspace$; the specialization to $\ki\=\k, \,
\kii\=1$ affords the relevant results for the harmonic oscillator
in the three `classical' Riemannian spaces of constant curvature
$({\bf S}^2_{\k},{\bf E}^2,{\bf H}^2_{\k})$, and further
specialization to $\k=0$ leads to the harmonic oscillator in ${\bf
E}^2$. The results also cover the harmonic oscillator on
Lorentzian configuration spaces.

After an introduction, we have divided this section into three
subsections: in the first part we consider the equivalent
one--dimensional problem and we draw some information and a
classification of the orbits previous to  obtaining  any
closed-form solution for the motion; in the second part we solve
explicitly the problem and we obtain closed expressions of the
orbits. The results can be obtained by using three different
methods: first by direct integration, second by obtaining the
$\CKspace$--dependent version of the Binet's equation, and third
by exploiting anew the superintegrability of the problem; this
last method is closer to the one usually employed in the Euclidean
oscillator, which leans on its separable character in Cartesian
coordinates, a property which is not shared by the Kepler
potential. The third subsection is devoted to the analysis and
classification of orbits, with the emphasis restricted to the
three `classical' Riemannian configuration spaces with nonzero
curvature $\k$. This third subsection also interprets the
trajectories obtained above as conics in curved spaces; of course
all the results reduce to well known Euclidean trajectories
---ellipses centred in the potential origin---, when we specialize
the parameters to their standard Euclidean values  $\ki\=0,
\kii\=1$. Our description of conics goes beyond the classical
papers on this topics published around 1900 (where the emphasis
was mainly projective, although some metrical aspects are also
discussed; see, e.g., \cite{Sto82} and references therein or
\cite{CooBook, KleinBook}). And, at any rate, conics in locally
Minkowskian were definitely not considered at all in any of these
works; probably \cite{Bi62} is the first paper dealing with this
subject.

Sec.\ IV, that has a more geometric character, gives some
information on the theory of conics in the general CK space
$\CKspace$, leading to the identification of the harmonic
oscillator orbits with conics for any value of the parameters
$\ki, \kii$ (this is, either for Riemannian as well as for
Lorentzian configuration spaces), and serves as a geometrical
counterpart and complement to the information already provided on
Sec.\ III. Additional details are provided mainly in the
Riemannian case; we plan to discuss the case of a Lorentzian
configuration space in more extension in a forthcoming paper.
Finally, in Sec.\ V we make some final comments.

We mention that most results here also hold (in a suitably reformulated way)
for the $n$-dimensional version of the harmonic oscillator, which is well
known to be superintegrable \cite{HBSS04, LoMR99}.


\section{Dynamics on $\CKspace$: Geodesic motion, Noether symmetries and
constants of motion}

We start by discussing some details on the motion of a particle in
a configuration space $\CKspace$ in the general CK case, where the
parameters $\ki, \kii$ may have any real value. This
$(\ki,\kii)$-dependent formalism contains {\it nine} essentially
different Cayley-Klein spaces, because by scaling the units of
length by $\lambda$ and of angle by $\alpha$, the values of $\ki$
and $\kii$ transform as $\ki\to\lambda^2 \ki$, $\kii\to\alpha^2
\kii$. Then without any loss of generality, any CK space can be
brought to its {\it standard form}, with either $\ki=1, 0, -1$ and
$\kii=1, 0, -1$. The standard form of any expression in the CK
formalism coincides with the result one would obtain by working
from the outset in a single space. Should we proceed this way,
however, the consideration of how details change when there is a
variation of curvature or signature type would require an
additional separate study. The distinctive trait in the CK
formalism is that the dependence on $\ki$ and $\kii$ is built-in,
and makes a further study of the limiting processes fully
redundant. For this reason we will keep the general form, with
explicit $\ki$ and $\kii$ in most of the paper, stressing when
required the specific properties holding after a specialization
for the values of $\ki, \kii$.

The presence of {\it two parameters} in the CK family of
two-dimensional spaces is related to the Cayley-Klein theory of
projective metrics, and underlies the length/angle duality which
is the residue of the general duality in projective geometry when
projective metrics are taken into account. It is also related to
the existence of two commuting involutions in their isometry Lie
algebras \cite{HeSa96}

Within the generic {\it standard} choices $\kii=\pm1$ for the
signature type, the CK family $\CKspace$  with $\kii\=1$ includes
the three `classical' Riemannian spaces with constant curvature
$\ki=\k$ and the CK family $\CKspace$ with $\kii\=-1$ includes the
three Lorentzian spaces with constant curvature (kinematically
interpretable as homogeneous space-times). In the non-generic case
$\kii\=0$ the CK spaces can be interpreted as the three
$\unomasuno $ non-relativistic space-times, which are limits of
the spaces with $\kii\!\neq\!0$; see \cite {HeOrSa00,OrSa02} and
references therein. These nine spaces can be conveniently
displayed in a Table; for more details see \cite{HeOrSa00,
ConformalHS02, ConformalCompactHS02}.

\begin{table}[h]
{\footnotesize
  \noindent
\caption{{The nine standard two-dimensional CK
spaces $\CKspace$.}}\label{ckho07:table:9CKGeometries}
\noindent\hfill
\begin{tabular}{llll}
\hline\\[-8pt]
  &\multicolumn{3}{c}{Measure of distance \& Sign of $\ki$}\\[2pt]
\cline{2-4}\\[-8pt]
Measure of angle\ &Elliptic&Parabolic&hyperbolic\\
\ \ \& Sign of $\kii$&$\ki=1$&$\ki=0$&$\ki=-1$\\[2pt]
\hline\\[-8pt]\hline\\[-8pt]
&Elliptic&Euclidean&hyperbolic\\
Elliptic $\kii=1$&${\bf S}^2$&${\bf E}^2$&${\bf H}^2$\\[2pt]
\hline\\[-8pt]
&Co-Euclidean&Galilean&Co-Minkowskian\\
&Oscillating NH & &Expanding NH\\
Parabolic $\kii=0$&${\bf ANH}^\unomasuno$&${\bf G}^\unomasuno$&${\bf 
NH}^\unomasuno$\\[2pt]
\hline\\[-8pt]
&Co-hyperbolic&Minkowskian&Doubly hyperbolic\\
&Anti-de Sitter& &De Sitter\\
hyperbolic $\kii=-1$&${\bf AdS}^\unomasuno$&${\bf 
M}^\unomasuno$&${\bf dS}^\unomasuno$\\[2pt]
\hline\\[-8pt]\hline\\[-8pt]
\end{tabular}\hfill}
\end{table}

On any general two-dimensional Riemannian $V^2$ or Lorentzian
space $L^2$, not necessarily of constant curvature, there are two
distinguished types of local coordinate systems, `geodesic
parallel' and `geo\-desic polar', that reduce to the familiar
Cartesian and polar coordinates on the Euclidean or Minkowskian
plane  \cite{doCa, Klin78} (see Appendix). The $\k$-dependent
Kepler problem in $V^2_\k$ was studied in \cite{CRS05} only in
polar coordinates but, since the Euclidean oscillator allows
separation also in parallel coordinates and this property is
shared for its `curved' version, we will use in this paper both
types of `geodesic' coordinates: polar $(r,\phi)$ and parallel
$(u,y)$.

\subsection{Polar coordinates}

The following expression, where $\ki, \kii$ are two real  parameters,
\begin{equation}
   ds^2 = d r^2 + \kii \CKSrSq\,d{\phi}^2 \,,
\end{equation}
represents, in polar coordinates $(r,\phi)$, the differential line
element  on the space $\CKspace$. When $\kii\=1$, these spaces
are, according to $\ki>, =, < 0$, the three {\it classical}
Riemannian spaces $V^2_{\ki}\equiv {\bf S}^2_{\ki},{\bf E}^2,{\bf
H}^2_{\ki}$ with constant curvature ${\ki}$. In the three standard
cases $\ki=1, 0, -1$ the metrics correspond respectively to the
{\it standard} sphere ${\bf S}^2$ $(\ki\=1,\kii\=1)$, Euclidean
plane ${\bf E}^2$ $(\ki\=0, \kii\=1)$, and hyperbolic or
Lobachevsky plane ${\bf H}^2$ $(\ki\=-1,\kii\=1)$:
$$
   \left.ds^2\right|_{{\bf S}^2} = d r^2 + (\sin^2 r)\,d{\phi}^2 \,,{\quad}
   \left.ds^2\right|_{{\bf E}^2} = d r^2 + r^2\,d{\phi}^2 \,,{\quad}
   \left.ds^2\right|_{{\bf H}^2} = d r^2 + (\sinh^2 r)\,d{\phi}^2\,,
$$

Likewise, when $\kii\=-1$ these spaces are the pseudo--Riemannian 2d
spaces $L^2_{\ki}$ with indefinite non-degenerate metric (hence Lorentzian)
of constant curvature $\ki$ and for the standard values $\ki=1, 0, -1$
the metrics reduces to:
$$
   \left.ds^2\right|_{{\bf AdS}^{\unomasuno}} = d r^2 - (\sin^2 
r)\,d{\phi}^2 \,,{\quad}
   \left.ds^2\right|_{{\bf M}^{\unomasuno}} = d r^2 - r^2\,d{\phi}^2 \,,{\quad}
   \left.ds^2\right|_{{\bf dS}^{\unomasuno}} = d r^2 - (\sinh^2 r)\,d{\phi}^2\,,
$$
which correspond to the three {\it standard} Lorentzian spaces
${\bf AdS}^{\unomasuno}, {\bf M}^{\unomasuno}, {\bf
dS}^{\unomasuno}$. Here polar coordinates only cover the region
with `time-like' separation from the origin;  unlike the
Riemannian case, ${\bf AdS}^{\unomasuno}$ and ${\bf
dS}^{\unomasuno}$ are related by a change of sign in the metric,
and are thus essentially the same space; this transformation is
conveyed by the change $r \leftrightarrow \ii r$, which
interchanges the regions with time-like and space-like separation
from the origin.

The three vector fields $X_{P_1}$, $X_{P_2}$, $X_{J}$, whose
coordinate expressions are given by:
\begin{eqnarray}
   X_{P_1} &\=& \CKCphi \,\fracpartial{r}
   - \frac{\CKSphi}{\CKTr} \,\fracpartial{\phi}   \,,\cr
   X_{P_2} &\=& \kii \CKSphi \,\fracpartial{r}
   + \frac{\CKCphi}{\CKTr}\,\fracpartial{\phi} \,,\cr
   X_J &\=& \fracpartial{\phi}  \,,
\label{ckho07:CKKillFCPol}
\end{eqnarray}
are  Killing vector fields of the metric, and are well defined for
any value of $\ki, \kii$ (even in the most degenerate CK space,
the Galilean or isotropic space where $\ki\=0, \kii\=0$). Each $X$
generates a one-parameter group of isometries of the metric and
altogether close on a Lie algebra denoted $\mathfrak{so}_{\ki,
\kii}(3)$:
\begin{equation}
[X_J,X_{P_1}]=-X_{P_2} \qquad [X_J,X_{P_2}]=\kii X_{P_1} \qquad
[X_{P_1},X_{P_2}]=-\ki X_J.
\label{ckho07:CKKillFCommRel}
\end{equation}

Of course, when $\ki, \kii$ are set to any particular values, all
these expressions give the pertinent ones for the corresponding
spaces; this is the trait in all the CK formalism. Notice that
only when $\ki\=0$ (Euclidean, Galilean and Minkowskian plane)
$X_{P_1}$ and $X_{P_2}$ commute. If we restrict to the family of
classical homogeneous Riemannian spaces $V_{\ki}^2$ with curvature
$\ki$, the corresponding Killing vector fields are given by
setting $\kii\=1$ in (\ref{ckho07:CKKillFCPol}), and then the
angular coordinate appears through the circular trigonometric
functions in the three spaces, where the radial coordinate appears
through the $\ki$-ones, which are either circular, parabolic or
hyperbolic according to the sign of $\ki$:
\begin{eqnarray}
      \left.X_{P_1}\right|_{V^2_\k} &=& (\cos{\phi})\,\fracpd{}{r} -
\Bigl(\frac{\CKCr}{\CKSr}\sin{\phi}\Bigr)\,\fracpd{}{\phi} \,,\cr
      \left.X_{P_1}\right|_{V^2_\k} &=& (\sin{\phi})\,\fracpd{}{r} +
\Bigl(\frac{\CKCr}{\CKSr}\cos{\phi}\Bigr)\,\fracpd{}{\phi} \,,\cr
      \left.X_{J} \right|_{V^2_\k}  &=& \fracpd{}{\phi} \,.    {\nonumber}
\end{eqnarray}

Moreover, the Lagrangian for a (free) particle moving in a configuration
space $\CKspace$ is given by the kinetic term invariant under the actions
of $X_{P_1}$, $X_{P_2}$, $X_{J}$ arising from the metric:
$$
  \CKLag_0(r, \phi, v_r, v_\phi) = T_{(\ki, \kii)}(r, \phi, v_r, v_\phi) =
  \smallonehalf\,\bigl(\,v_r^2 + \kii\CKSrSq v_\phi^2\,\bigr) \,,
$$
A general {\it natural\/} Lagrangian in $\CKspace$ (kinetic minus
potential term) has the following form
$$
  \CKLag(r, \phi, v_r, v_\phi) = \smallonehalf\,\bigl(\,v_r^2 + \kii \CKSrSq
  v_{\phi}^2\,\bigr) -  \CKPot(r,\phi)  \,,
$$
in such a way that for $\ki\=0, \kii\=1$ we recover a standard
Euclidean system
$$
  L(r, \phi, v_r, v_\phi) = \lim_{\ki\to 0, \kii\=1}\,\CKLag(r, \phi, 
v_r, v_\phi)
  = \smallonehalf\,(v_r^2 + r^2 v_{\phi}^2) - V(r,\phi)
  \,,{\quad}   V(r,\phi) = \lim_{\ki\to 0, \kii\=1}\, \CKPot(r,\phi) \,.
$$
In some particular cases a Lagrangian system can possess the
Killing vector fields $X_{P_1}$, $X_{P_2}$, or $X_{J}$ (or any
linear combination of them) as exact Noether symmetries. If we
denote by $X^t$ the natural tangent lift to the tangent bundle
(velocity phase space) of the vector field $X$  and by $\theta_L$
the Cartan semibasic one-form \cite{MaS85}
$$
  \theta_L = \fracpd{L}{v_r}\, dr + \fracpd{L}{v_\phi}\, d\phi
   =  v_r\,dr + \kii \CKSrSq v_{\phi}\,d\phi \,,
$$
then the {\it basic} cases with exact Noether symmetries are the following:

\begin{enumerate}
\item  If the potential $\CKPot(r, \phi)$ is invariant under 
$X_{P_1}$, then $\CKPot(r, \phi)$ should depend on $(r, \phi)$ only 
through an arbitrary function of the single variable 
$z_2\equiv\CKSr\CKSphi$ and then
$$
  P_1 = i(X_{P_1}^t)\,\theta_L =
  (\CKCphi)\,v_r - \kii (\CKCr \CKSr \CKSphi)\,v_{\phi}
$$
is a constant of motion.

\item  If the potential $\CKPot(r, \phi)$ is invariant under 
$X_{P_2}$, then $\CKPot(r, \phi)$ should be an arbitrary function of 
the single variable $z_1\equiv\CKSr\CKCphi$ only, and then
$$
   P_2 = i(X_{P_2}^t)\,\theta_L =
   \kii(\CKSphi)\,v_r + \kii(\CKCr \CKSrSq\CKCphi)\,v_{\phi}
$$
is a constant of motion.

\item If the potential $\CKPot(r, \phi)$ is invariant under
$X_{J}$, then $\CKPot(r, \phi)$ should be an arbitrary function of
the single variable $r$ only ($\CKPot$ is a central potential) and
the constant of motion is:
$$
   J = i(X_J^t)\,\theta_L = \kii \CKSrSq\,v_{\phi}\,.
$$
\end{enumerate}

Several remarks are pertinent. First, the radial dependence in the 
momenta $P_1, P_2, J$ appears through $\ki$-trigonometric functions, 
and hence is sensitive to the curvature; the angular dependence is 
carried through $\kii$-trigonometric functions and in the three {\it 
classical} ($\kii\=1$) Riemannian spaces it appears through 
$\cos{\phi}$ or $\sin{\phi}$, irrespectively of the curvature.
Second, the quantities $P_1$, $P_2$, $J$, could be considered as the 
ordinary linear momenta and angular momentum of a particle moving in 
the configuration space $\CKspace$. In terms of these, the kinetic 
energy $T_{(\ki, \kii)}$ can be rewritten as follows (it is the 
Casimir of the isometry algebra, \cite{HeSa97})
$$
   T_{(\ki, \kii)} = \frac12\,\frac{
         \kii P_1^2 + P_2^2 + \ki\,J^2}{\kii} \,,
$$
showing that, on spaces of (constant) non-zero curvature $\ki$, the
angular momentum has a contribution to the kinetic energy of the
system, proportional to the curvature $\ki$. And third, the new 
quantities $P_2$ and $J$ vanish {\em identically} when $\kii\=0$. 
This is linked to the singular character of the corresponding 
Lagrangian, as the metric is degenerate when $\kii\=0$. This singular 
case is however not {\it generically} singular, but only a very 
special limit of a regular system, and one may expect the geodesic 
motion to have precisely three non-trivial constants of motion linear 
in the velocities. When working in the general CK scheme, where we 
want to cover all CK spaces $\CKspace$ (even when $\kii\=0$) this 
suggests to consider, instead of $P_1, P_2, J$, the quantities 
defined as
\begin{equation}
\CKP_1:=P_1, \qquad
\CKP_2:=\frac{P_2}{\kii},\qquad
\CKJ:=\frac{J}{\kii},
\end{equation}
which will be called the {\it CK Noether momenta}; when
$\kii\!\neq\!0$ these  are essentially equivalent to $P_1, P_2, J$
but $\CKP_1, \CKP_2, \CKJ$ are to be preferred because they remain
non vanishing even in the limit $\kii\to0$. These will be the
momenta used in the rest of the paper; we remark that in the {\it
classical}  Riemannian spaces of constant curvature $\ki$ the
three Noether momenta {\it coincide} with $P_1, P_2, J$. In
$\CKspace$ the CK Noether momenta are:
\begin{eqnarray}
   {\CKP_1} &=& \CKCphi\, v_{r} - \kii\CKCr\CKSr\CKSphi\, v_{\phi}\,,\\[2pt]
  {\CKP_2} &=& \CKSphi\, v_{r} + \CKCr\CKSr\CKCphi\, v_{\phi}\,,\\[2pt]
   \CKJ &=& \CKSrSq\, v_{\phi}\,.
\label{ckho07:CKNoeMomCPol}
\end{eqnarray}

In terms of these Noether momenta,  the kinetic energy is well
defined for all CK spaces, contains always a term $\CKP_1^2$ and
is given by:
\begin{equation}
T_{(\ki, \kii)}=
\smallonehalf \left(\CKP_1^2 + \kii\CKP_2^2 + \ki\kii\CKJ^2\right).
\label{ckho07:CKKinEneNM}
\end{equation}

\subsection{Parallel coordinates}

The element of arc length $ds^2$ in the space $\CKspace$ is given
in parallel coordinates $(u,y)$ by
\begin{equation}
ds^2 = \CKCySq \, du^2 + \kii \, dy^2\,,
\end{equation}
and in the three particular classical Riemannian standard cases it
reduces to
$$
   \left.ds^2\right|_{{\bf S}^2} =  (\cos^2 y)\,d u^2 + d y^2 \,,{\quad}
   \left.ds^2\right|_{{\bf E}^2} =  d u^2 + d y^2 \,,{\quad}
   \left.ds^2\right|_{{\bf H}^2} = (\cosh^2 y)\,d u^2 + d y^2\,.
$$

The Lagrangian of a free particle in $\CKspace$ has a kinetic term
corresponding to the metric:
$$
   \CKLag(u, y; v_u, v_y) = T_{(\ki, \kii)}(u, y; v_u, v_y) =
   \smallonehalf\,\bigl(\,\CKCySq\,v_{u}^2 + \kii v_{y}^2\,\bigr)\,.
$$
In these coordinates the three Killing vector fields closing the Lie
algebra $\mathfrak{so}_{\ki, \kii}(3)$ are
\begin{eqnarray}
   X_{P_1} &\=& \fracpartial{u}  \,,\cr
   X_{P_2} &\=& \ki\kii \CKSu \CKTy\,\fracpartial{u} + 
\CKCu\,\fracpartial{y}\,,\cr
   X_J &\=& -\kii\CKCu\CKTy\,\fracpartial{u} + \CKSu\,\fracpartial{y}\,,
\label{ckho07:CKKillFCPar1}
\end{eqnarray}
and the associated momenta
\begin{eqnarray}
   P_1 &\=& i(X_{P_1}^t)\,\theta_L
   = \CKCySq\,v_{u}   {\nonumber}\,,\\[2pt]
   P_2 &\=& i(X_{P_2}^t)\,\theta_L
   = \ki\kii\,\CKSu \CKCy \CKSy \,v_{u} + \kii \CKCu \,v_{y}  \,,{\nonumber}\,,
   \\[2pt]
   J   &\=& iX_{J}^t)\,\theta_L
   = - \kii\CKCu  \CKCy \CKSy \,v_{u} + \kii \CKSu \,v_{y} \,,{\nonumber}
\end{eqnarray}
lead to the CK Noether momenta:
\begin{eqnarray}
   {\CKP_1} &\=& \CKCySq  \,v_{u}\,,\\[1pt]
  {\CKP_2} &\=& \ki \CKSu \CKSy \CKCy  \,v_{u} + \CKCu  \,v_{y}\,,\\[1pt]
   \CKJ &\=& -\CKCu \CKSy \CKCy  \,v_{u} + \CKSu  \,v_{y}\,.
\label{ckho07:CKNoeMomCPar1}
\end{eqnarray}

Notice how the expressions  for the CK momenta specialize in the Euclidean
case:
$$
\left. \CKP_1 \right|_{{\bf E}^2} =  v_{u} \quad
\left. \CKP_2 \right|_{{\bf E}^2} =  v_{y} \quad
\left. \CKJ \right|_{{\bf E}^2} = u  \,v_{y} - y \, v_{u}.
$$
The general CK expressions can be looked at as a two-parameter
deformation of the Galilean ones $\ki\=0,\,\kii\=0$, governed by
the two constants $\ki, \kii$; the Euclidean case is {\it not} the
natural comparison standard in the deformation, as one of the
constants is already non-vanishing for ${\bf E}^2$.

A potential $\CKPot$, now expressed as a function of $(u, y)$,
turns out to be invariant under $X_{P_1}$ if $\CKPot$ is an
arbitrary function of the single variable $y$ only; this result is
simply the translation to parallel coordinates of the previous
result in polar coordinates ---as consequence of the relation
$\CKSy = \CKSr\CKSphi$---, so that the auxiliary variable $z_2$
turns out to be precisely $\CKSy$. Similar results describe the
general form, in coordinates $(u, y)$ of a potential  invariant
under $X_{P_2}$ or under $X_{J}$.

We close this section with a comment on the expressions obtained
for the three vector fields, $X_{P_1}$, $X_{P_2}$, $X_J$ (for
$\kii\=1$ this point was discussed in \cite{RaSa03}). According to
the straightening-out theorem \cite{AbMaRa}, a vector field $X$ on
a $n$-manifold $V$ always admits a local coordinate system
$\{x_1,\dots,x_n\}$ in an appropriate neighbourhood of a regular
point  $X(m)\ne 0$, $m\in M$, such that then it becomes $Y=\sum_k
c_k (\partial{}/\partial{x_k})$, with $c_{k_0}=1$, $c_k=0$ for
$k\ne k_0$. We recall that in polar coordinates $X_J$ is given by
$X_J=\partial{}/\partial{\phi}$ and now we have obtained that in
parallel $(u,y)$ coordinates $X_{P_1}$ takes the form
$X_{P_1}=\partial{}/\partial{u}$. Similarly, in the complementary
`orthogonal' parallel system $(x,v)$, we obtain
$X_{P_2}=\partial{}/\partial{v}$ (this parallel system is not used
in this paper but is discussed in the Appendix). So, these three
coordinate systems, $(r,\phi)$, $(u,y)$, and $(x,v)$, are the
three appropriated systems (via the straightening-out theorem)
providing the `straight' expressions of $X_J$, $X_{P_1}$, and
$X_{P_2}$, respectively.


\section{The harmonic oscillator on $\CKspace$ }

The following Lagrangian in the CK space $\CKspace$ with curvature
$\ki$ and signature type $\kii$
\begin{equation}
\CKLag(r, \phi, v_r, v_\phi)= \smallonehalf\, \big( v_r^2 + \kii 
\CKSrSq\,v_{\phi}^2 \big) - \CKPot_{HO}(r),
\qquad
\CKPot_{HO} =\smallonehalf \omega_0^2\, \CKTrSq\,,
\end{equation}
represents the `harmonic oscillator' in the space $\CKspace$
\cite{RaS02b,RaSa03}; the potential $\CKPot_{HO}(r)$ is `central'
in the sense it depends on the radial coordinate only; this
dependence involves the label $\ki$ (but not $\kii$) and reduces
to
$$
   \left.\CKPot_{HO}\right|_{\ki\=1} =  \smallonehalf\, 
\omega_0^2\,\tan^2\!r  \,,{\quad}
   \left.\CKPot_{HO}\right|_{\ki\=0}= V = \smallonehalf\, 
\omega_0^2\,r^2  \,,{\quad}
   \left.\CKPot_{HO}\right|_{\ki\=-1} = \smallonehalf\, 
\omega_0^2\,\tanh^2\!r \,,
$$
which are the three harmonic oscillator potentials in the three
classical standard Riemannian spaces ${\bf S}^2, {\bf E}^2, {\bf
H}^2$. The Euclidean function $V(r)$ appears in this formalism as
making a separation between two different behaviours (see Figure
1). In the sphere this potential was considered by Liebmann
\cite{Lieb} (1905 edition), and later on by Higgs \cite{Hi79} and
Leemon \cite{Le79}.

Recall $r$ denotes the distance to the origin point computed in
the intrinsic metric on the CK space. Thus in the classical
Riemannian case ($V^2_{\ki}, \kii\=1$), the potential has a zero
(minimum) value at the origin and starts growing  quadratically
with the distance to the origin point, as implied by the
approximation $\CKTr \approx r$ around $r=0$, which holds for all
values of $\ki$. In the flat case the potential grows
quadratically with any $r$ and approaches an infinite value only
when $r\to\infty$. When the curvature is non-zero, the behaviour
differs in a way depending on the curvature sign.  When $\ki$ is
positive the potential grows faster and tends to infinity at a
finite value $r=\frac{\pi}{2\sqrt{\ki}}$, this is, on the sphere
`equator' (with the origin taken as the pole); the harmonic
oscillator on the sphere splits the configuration space into two
halves by an infinite potential wall on the equator, so the
spherical harmonic oscillator has two antipodal centres. In the
negative curvature case, the potential grows slower than in the
flat case, and as $r\to\infty$ approaches a plateau, with a
(positive) finite height $\CKPot_\infty:=\omega_0^2/(-2\ki)$.

The motion in this potential  is {\it superintegrable} in all CK
spaces $\CKspace$ since, in addition to the angular momentum
$\CKJ$, this system is endowed with the following quadratic
constants of the motion
\begin{equation}
\begin{tabular}{ll}
$I_{\CKJ^2}=\CKJ^2$\,,\\[2pt]
$I_{\CKP_1^2}=\CKP_1^2 + \CKW_{11}(r,\phi)$,
            \qquad & $\CKW_{11}(r,\phi) = \omega_0^2 \CKTrSq 
\CKCphiSq$\,,\\[2pt]
$I_{\CKP_2^2}=\CKP_2^2 + \CKW_{22}(r,\phi)$
            \qquad & $\CKW_{22}(r,\phi) = \omega_0^2 \CKTrSq \CKSphiSq 
$\,,\\[2pt]
$I_{\CKP_1\CKP_2}=\CKP_1\CKP_2 + \CKW_{12}(r,\phi)$
            \qquad & $\CKW_{12}(r,\phi) =\omega_0^2 \CKTrSq \CKCphi\CKSphi $\,.
\end{tabular}
\label{ckho07:CKHOIconstCPol}
\end{equation}
Remark that $\CKW_{11}(r,\phi), \CKW_{22}(r,\phi),
\CKW_{12}(r,\phi)$ are well defined for any CK space, and they do
not vanish identically in none of them.

The `energy' of the motion can be written as:
\begin{equation}
I_E = \smallonehalf (I_{\CKP_1^2} + \kii I_{\CKP_2^2} +
          \ki\kii I_{\CKJ^2})
          = \smallonehalf ({\CKP_1^2} + \kii {\CKP_2^2} +
          \ki\kii {\CKJ^2}) + \smallonehalf \omega_0^2 \CKTrSq\,,
\label{ckho07:CKHOIERelIType}
\end{equation}
reducing on ${\bf E}^2$ to the known Euclidean expression. In the
general CK space $\CKspace$ the kinetic energy is no longer given
by (one half of) the `norm' of the `momentum vector' ${\CKP_1^2} +
\kii {\CKP_2^2}$ but contains as an {\it extra  contribution} the
square of the angular momentum, proportional to the curvature
$\ki$ and thus disappearing in the flat case $\ki\=0$.

The four integrals of motion in (\ref{ckho07:CKHOIconstCPol})
cannot be functionally independent; indeed they satisfy the
relation:
$$
I_{\CKP_1^2}I_{\CKP_2^2} - (I_{\CKP_1\CKP_2})^2 =
          \omega_0^2 I_{\CKJ^2}\,.
$$

Taken altogether the constants $I_{\CKP_1^2}, I_{\CKP_2^2}, I_{\CKP_1\CKP_2}$
are the components of a (symmetric) tensor under the `rotation subgroup'
$SO_{\kii}(2)$ in any space $\CKspace$
\begin{equation}
\left(\begin{array}{cc}
\CKF_{11}&\CKF_{12}\cr
\CKF_{21}&\CKF_{22}
\end{array}\right)
=
\left(\begin{array}{cc}
I_{\CKP_1^2}
&
I_{\CKP_1\CKP_2}
\cr
I_{\CKP_1\CKP_2}
&
I_{\CKP_2^2}
\end{array}\right)
=
\left(\begin{array}{cc}
\CKP_1^2 + \CKW_{11}(q_1,q_2)
&
\CKP_1\CKP_2 + \CKW_{12}(q_1,q_2)
\cr
\CKP_1\CKP_2 + \CKW_{12}(q_1,q_2)
&
\CKP_2^2 + \CKW_{22}(q_1,q_2)
\end{array}\right)
\end{equation}
Thus the essential property of the Euclidean harmonic oscillator,
to have a tensor constant of motion (the so called Fradkin tensor
\cite{Fr65, JaH40}), survives for the `curved' harmonic oscillator
in any CK space $\CKspace$. This tensor, looked at as the general
CK form of the Fradkin tensor, contains also a complete set of
functionally independent constants of motion, because $\CKJ^2 =
I_{\CKJ^2}$ is related to the determinant of the Fradkin tensor in
a `universal' way, with a relation explicitly independent of $\ki,
\kii$ :
\begin{equation}
\det(F) = \omega_0^2 I_{\CKJ^2}
\label{ckho07:CKHOdetF}
\end{equation}
Another possible choice for three functionally independent
constants is the set $\{I_{\CKP_1^2}, I_{\CKP_2^2}, I_{\CKJ^2}\}$.

\subsection{The classification of orbits and the equivalent
one-dimensional problem}

It is very convenient to introduce in the study of central
potentials $V(r)$ on the Euclidean space a one-dimensional
effective potential $V^{\rm eff}$, which governs the radial motion
after elimination of the ignorable angular coordinate. This
procedure allows us to obtain a classification of orbits before
the integration of the equations \cite{Go80}. Given a central
potential $\CKPot$ on the CK spaces $\CKspace$ one can proceed
similarly and this leads to the one-dimensional `effective'
potential $\CKPoteff(r)$:
$$
   \CKPoteff(r) = \CKPot(r) + \frac{\kii \CKJ^2}{2 \CKSrSq} \,,
$$
where the extra term plays the role of the `centrifugal barrier
potential', reducing as it should be to ${\CKJ^2}/{(2 r^2)}$ in
the standard Euclidean case. Therefore we can classify the orbits
and obtain some additional information for the harmonic oscillator
motion in any CK space $\CKspace$ simply by analyzing the
effective potential for the harmonic oscillator motion:
\begin{equation}
   \CKPoteff(r) =
\smallonehalf\, \omega_0^2\,\CKTrSq + \frac{\kii \CKJ^2}{\,2\CKSrSq} =
\smallonehalf\, \omega_0^2\,\CKTrSq + \frac{\kii \CKJ^2}{\,2\CKTrSq} 
+\smallonehalf \ki\kii\CKJ^2
\,,
\end{equation}
which in the three particular one-dimensional problems associated
to the standard ($\ki=1,0,-1; \kii\=1$) sphere ${\bf S}^2$,
Euclidean plane ${\bf E}^2$, and Lobachevsky plane ${\bf H}^2$
reduces respectively to
\begin{eqnarray}
\left.\CKPoteff_{HO}\right|_{{\bf S}^2} &=&
   \smallonehalf\, \omega_0^2\,\tan^2 r + \frac{J^2}{\,2\sin^2 r} =
   \smallonehalf\, \omega_0^2\,\tan^2 r + \frac{J^2}{\,2\tan^2 r} + 
\smallonehalf J^2 \,,\cr
\left.\CKPoteff_{HO}\right|_{{\bf E}^2} &=&
   \smallonehalf\, \omega_0^2\,r^2  +  \frac{J ^2}{\,2 r^2} \,,\cr
\left.\CKPoteff_{HO}\right|_{{\bf H}^2} &=&
    \smallonehalf\, \omega_0^2\,\tanh^2 r + \frac{J ^2}{\,2\sinh^2 
r}=\smallonehalf\, \omega_0^2\,\tanh^2 r + \frac{J ^2}{\,2\tanh^2 r}- 
\smallonehalf J^2 {\nonumber}
\end{eqnarray}

The standard Euclidean case $\ki\=0, \kii\=1$ needs no comment. We
discuss in some detail the Riemannian cases, allowing any  nonzero
(positive) values for the curvature $\ki\!\neq\!0$ and signature
type $\kii\>0$. The discussion is made within the generic case
$\CKJ\neq0$, and we will assume $\CKJ>0$ by considering if
necessary the reversed motion along the same geometric orbit; for
$\CKJ=0$ the motion is actually one-dimensional. From now on we
also omit the label {HO} in the potential when it is clear from
the context.

\bigskip

\noindent{(1)} Spherical case: Analysis of the potential
$\CKPoteff$ for $\ki\>0, \kii\>0$.

This corresponds to motion in a sphere, and the natural range for
the $r$ coordinate is the interval $(0, \pi/(\sqrt{\ki})$, which
is the span of $r$ along half a geodesic (half a sphere's large
circle). Provided $\omega_0$ is real as implicitly assumed (the
constant in $\CKPot_{HO}$ is assumed to be positive) the effective
potential is always positive, $\CKPoteff(r)>0$, satisfies the
following limits in the boundaries
$$
   \lim_{r\to 0} \CKPoteff(r) = + \infty \,,\quad
   \lim_{r\to \pi/(2\sqrt{\ki})} \CKPoteff(r) = + \infty \,,
$$
and it has a minimum, whose value we denote $E(\CKJ)$, at the
point $r_m$ given by $\CKtSq(r_m) = \sqrt{\kii} \CKJ/\omega_0$
$$
E(\CKJ):= \CKPoteff(r_m)=\sqrt{\kii} \omega_0 \CKJ + \smallonehalf 
\ki\kii \CKJ^2\,.
$$
This equivalent potential represents therefore, an asymmetrical
well on the `$r$ line', with two barriers of infinite height at
$r=0$ and $r=\pi/(2\sqrt{\ki})$, and one single minimum placed in
between (Figure 2). Thus, for a fixed value of $\CKJ$ the
situation is as follows:  there is not any possible motion for
energies $E<E(\CKJ)$, there is the (unique)  motion $r(t)=r_m$
when $E=E(\CKJ)$, and for all energy values $E>E(\CKJ)$ the motion
of the radial coordinate consists of non-linear one--dimensional
oscillations between the two `radial' turning points. On the
sphere, the motion with a fixed value of $\CKJ$ and the minimum
compatible energy $E=E(\CKJ)$ is a circular motion, on the circle
centred in the origin with radius $r=r_m$. The trajectories with
$E(\CKJ)<E$ lie in a spherical annulus which always contains the
circle $r=r_m$; this annulus grows to cover all the hemisphere
when $E\to\infty$.

All these expressions depend on $\ki, \kii$ and reduce to the well
known expressions for the Euclidean oscillator when $\ki\=0,
\kii\=1$; notice  that the full Euclidean plane appears as the
limit $\ki\to0$ of a single hemisphere, with the sphere's
`equator' going to the Euclidean infinity.

\bigskip
\noindent{(2)} Hyperbolic case: Analysis of the potential $\CKPoteff$ for
$\ki\<0, \kii\>0$.

This is the hyperbolic Lobachevsky plane case, and here $r\in(0,
\infty)$. Provided again $\omega_0$ is real as implicitly assumed,
the effective potential is always positive, $\CKPoteff(r)>0$, and
has a minimum at  $r=r_m$ whenever $\CKt(r_m) = \sqrt{\kii}
\CKJ/\omega_0$. It satisfies the following limits in the
boundaries:
$$
   \lim_{r\to 0} \CKPoteff(r) = + \infty \,,\quad
   \lim_{r\to\,+\,\infty} \CKPoteff(r) = \frac{\omega_0^2}{-2\ki}\,,
$$
thus introducing into the problem a new energy scale, to be
denoted $E_\infty$
$$
E_\infty:= \frac{\omega_0^2}{-2\ki};
$$
(remark this scale could be also defined in the sphere case, where
its value will anyhow fall outside of the physically allowed range
of energies $[E(\CKJ), \infty]$). The value of the potential at
the minimum is given again by $E(\CKJ):=\CKPoteff(r_m)$, but here
the energy scale $E_\infty$ is placed above $E(\CKJ)$:
$E(\CKJ)<E_\infty$

When $\ki\<0$ the (absolute) values of the hyperbolic type tangent
$\CKTr$ are bounded by $\frac{1}{\sqrt{-\ki}}$; thus depending on
the values of $\CKJ$ there are two possible generic situations,
according to whether $\CKJ$ is  smaller or larger than an {\it
angular momentum scale}
$$
\CKJ_\infty:=\frac{\omega_0}{\sqrt{\kii}(-\ki)}
$$

\begin{enumerate}
\item{}  If $\CKJ<\CKJ_\infty$, this is if $\sqrt{\kii}
\CKJ/\omega_0<\frac{1}{-\ki}$ the function $\CKPoteff(r)$ has a
real minimum at the $r_m$ given by $\CKt(r_m) = \sqrt{\kii}
\CKJ/\omega_0$, with value $\CKPoteff(r_m)=\sqrt{\kii} \omega_0
\CKJ + \smallonehalf \ki\kii \CKJ^2$. Motions of the radial
coordinate with the given value of $\CKJ$ will only happen with
energies $E\geq E(\CKJ)$. Within this range of energies there are
two possibilities. For $E(\CKJ) < E < E_\infty$ the radial
coordinate is bounded, corresponding to a motion in the hyperbolic
plane in an annular region bounded by two circles (with radius
corresponding to the two turning points in the effective
one-dimensional radial potential); this annulus reduces to a
circle of radius $r=r_m$ for $E=E(\CKJ)$. When $E \to E_\infty$
while $E<E_\infty$  the exterior turning point approaches to
infinity, reaches it for $E=E_\infty$ and then the exterior
turning point disappears. Therefore for $E_\infty < E$ the motion
of the radial coordinate is unbounded, corresponding to motions in
the hyperbolic plane which stays outside of a geodesic circle
centred at the origin, whose radius corresponds to the still
existing interior turning point.

\item{} If $\CKJ\ge\CKJ_\infty$, this is if
$\sqrt{\kii}\CKJ/\omega_0{\ge}\frac{1}{\sqrt{-\ki}}$ the function
$\CKPoteff(r)$ has not any minimum, and decreases monotonically
from the potential barrier with infinite height at the origin
$r=0$ to the asymptotic plateau when $r\to\infty$; only energies
satisfying $E\ge E_\infty=\omega_0^2 / (-2\ki)$ will be allowed
here and all the trajectories will be unbounded (scattering) open
curves.
\end{enumerate}

These two possible behaviours and its separating case are
represented in Figure 3.

\bigskip
\noindent{(3)} Lorentzian case: Analysis of the potential
$\CKPoteff$ for $\kii\<0$.

A complete discussion of motion in a Lorentzian configuration
space will not be done here. We simply mention some traits. First,
the `centrifugal barrier' comes explicitly with the `opposite'
sign to the Riemannian one. Second, if $r$ is the distance to the
origin, the square $r^2$ can have any sign in a Lorentzian space
(with $r=0$ on the isotropes through the origin), and thus
formally $r$ will be real and positive on the region with
time-like separation from the origin, but pure imaginary in the
region with space-like separation. As the harmonic oscillator
potential depends on $\CKTrSq$, the potential will always be real,
yet it has the two signs on the two regions with time-like and
space-like separation from the origin. Hence the harmonic
oscillator potential has no isolated minima at any proper point,
and all points on the isotropes through the origin are extremal,
degenerate saddle points for the potential function. The possible
extremal values for the effective potential are still given by the
equation $\CKt(r_m) = \sqrt{\kii} \CKJ/\omega_0$. Because of the
inexistence of isolated minima for the potential, there is no
reason to require a positive $\omega_0^2$. If we accept a negative
sign for the constant $\frac12 \omega_0^2$, (i.e., purely
imaginary values for $\omega_0$, the products $\omega_0
\sqrt{\kii}$ and the quotients $\sqrt{\kii}/\omega_0$ would turn
real. In spite of these seemingly strange properties, the
Lorentzian harmonic oscillator is superintegrable, and thus
completely explicit solutions for its motion can be explicitly
given, irrespective of any possible  physical meanings which will
be very different from the familiar oscillator. We just remind
that the `inverted' or repulsive Euclidean harmonic oscillator
(with $\omega_0^2<0$) is also superintegrable, yet it only has
scattering motions along centred hyperbolas, which are also conics
with a centre at the origin; hence the complete set of conics with
centre at the origin appear as the complete set of orbits in the
harmonic oscillator only if we allow general, unrestricted,
values, for the strength constants.

\subsection{Determination of the orbits of the harmonic oscillator
in $\CKspace$}

\subsubsection{Method I: Direct Integration for a central potential}

We have previously obtained, making use of the conservation of the
total energy $E$ and of the angular momentum $\CKJ$ two expressions for
$\dot{r}$ and $\dot{\phi}$. Eliminating $t$ between both equations
we have
$$
   d\phi = \frac{J\,dr}{\CKSrSq\,\sqrt{\,R(r)\,}}
   \,,\qquad
   R(r) = 2 \,\bigg(\! \,E - \CKPot(r) - \frac{\kii \CKJ^2}{2\, 
\CKSrSq}\,\!\bigg)\,.
$$
After the change of variable $r \to \uRegPar$ with
$\uRegPar = 1/\CKTr$,  $dr = - \CKSrSq\,d\uRegPar$, this becomes
$$
   d\phi = -\,\frac{d\uRegPar}{\sqrt{\widehat{R}(\uRegPar)}} \,,\qquad
   \widehat{R}(\uRegPar) = \frac{2\,E}{\CKJ^2} - 
\frac{2\,\CKPot(\uRegPar)}{\CKJ ^2} - \kii (\uRegPar^2+\ki)\,.
$$
All this holds for a general $\CKPot$. Particularizing
for the harmonic oscillator in $\CKspace$, $\CKPot(\uRegPar) =
\smallonehalf \omega_0^2/\uRegPar^2$
$$
   \widehat{R}(\uRegPar) =\, \frac{2 E_P}{\CKJ ^2} -\,
   \frac{\omega_0^2}{\CKJ ^2}\frac{1}{\uRegPar^2} -\kii \uRegPar^2 \,,
$$
where we have used the notation $E_P = E - \smallonehalf
\,\ki\kii\, \CKJ^2$. An integration leads to
$$
\phi-\phi_0 = -\int \frac{d\uRegPar}{\sqrt{{\widehat{R}}(\uRegPar)}}\,,
$$
and the change $\chi = \uRegPar^2$
\begin{eqnarray}
   \phi - \phi_0 = -\, \frac12 \int 
\frac{d\chi}{\sqrt{{\widehat{R'}}(\chi)}}   \,,\qquad
  {\widehat{R'}}(\chi) = -\frac{\omega_0^2}{\CKJ ^2} + \frac{2 
E_P}{\CKJ ^2} \chi - \kii \chi^2 \,,
{\nonumber}
\end{eqnarray}
makes the integration elementary. In this way we arrive at the
general solution of the form
\begin{equation}
   \frac{1}{\CKTrSq} = D  -\,G\,\CKcc(2(\phi-\phi_0))\,,\qquad
\end{equation}
where of course the constant $\phi_0$ is trivial in the sense that
orbits with different values for $\phi_0$ are permuted by
isometries of the space, as corresponds to the central nature of
the potential. The two constants $D, G$ should be related to the
values of the energy and angular momentum $E, \CKJ$ as it follows
from the integration:
\begin{equation}
D = \frac{E_P}{\kii \CKJ^2}, \qquad
G = \frac{1}{\kii \CKJ^2}\sqrt{\,E_P^2 - \kii \omega_0^2\, \CKJ ^2\,},
\qquad
E_P=E-\smallonehalf \ki \kii \CKJ^2 \,.
\end{equation}

Hence the particular orbit corresponding to energy $E$ and angular
momentum $\CKJ$ is:
\begin{equation}
   \frac{1}{\CKTrSq} = \frac{E_P}{\kii \CKJ^2}
-\,\frac{1}{\kii \CKJ^2}\sqrt{\,E_P^2 - \kii \omega_0^2\, \CKJ ^2\,}
   \,\CKcc(2(\phi-\phi_0))\,,\qquad
\end{equation}
which can be also rewritten introducing two new constants $A, B$ as:
\begin{equation}
   \frac{1}{\CKTrSq} = \frac{\CKccSq(\phi-\phi_0)}{A^2} + 
\frac{\CKssSq(\phi-\phi_0)}{B^2}\,,
   \label{ckho07:CKHOOrbitSPCPol}
\end{equation}
where
\begin{equation}
   \frac{1}{A^2} = D - G  \qquad   \frac{1}{B^2} = \kii (D + G),
\end{equation}
are related to energy and angular momentum by the relations (to be
stated later in a simpler form)

\begin{equation}
   \frac{1}{A^2} =
  \frac{E_P -\,\sqrt{\,E_P^2 - \kii \omega_0^2\, \CKJ ^2}\,}
  {\kii \CKJ^2},
   \qquad
   \frac{1}{B^2} =
  \frac{E_P +\,\sqrt{\,E_P^2 - \kii \omega_0^2\, \CKJ ^2}\,}{\kii \CKJ^2}.
\end{equation}
Note that in the Riemannian case (when $\kii\>0$), then $0<G<D$,
and therefore both $D-G$ and $D+G$ are positive, thus making well
adapted the notation we have chosen for $A^2, B^2$; in the
Lorentzian case things behave differently.

All these equations apply for any CK space. In particular, in the
three classical {\it standard} Riemannian spaces  $({\bf S}^2,{\bf
E}^2,{\bf H}^2)$ with curvature $\ki\=1, 0, -1; \kii\=1$, and
chosing the origin for $\phi$ so that $\phi_0=0$, the orbit
equations are:
\begin{eqnarray}
\begin{tabular}{ll}
\hskip7pt$\displaystyle\frac{1}{\tan^2r} = \frac{1}{A^2}\,\cos^2\phi
+  \frac{1}{B^2}\,\sin^2\phi$ & in ${\bf S}^2$ \,,{\nonumber}\\[8pt]
\hskip22pt$\displaystyle\frac{1}{r^2}  =  \frac{1}{A^2}\,\cos^2\phi
           +  \frac{1}{B^2}\,\sin^2\phi$ & in ${\bf E}^2$ \,,{\nonumber}\\[8pt]
$\displaystyle\frac{1}{\tanh^2r} = \frac{1}{A^2}\,\cos^2\phi
           +  \frac{1}{B^2}\,\sin^2\phi$ & in ${\bf H}^2$ \,,{\nonumber}
\end{tabular}
\end{eqnarray}

Let us close this direct integration approach with three
observations. First, it turns out that in any of the nine CK
spaces, the HO orbit is always a {\it conic} with centre at the
potential origin, where `conic' has to be taken in a metric sense,
relative to the intrinsic metric in each space. This follows from
the geometrical study to be presented in Section 4. In the
Euclidean case the quantities $A$ and $B$ are directly the ellipse
semiaxes; in the case $\ki\!\neq\!0$ the semiaxes (understood in
terms of the intrinsic metric) are related to $A, B$ by some
relations which will involve $\ki, \kii$. Second, the connection
between the coefficients $D, G$ or $A, B$ and the energy and
momentum, when reexpressed in terms of the quantity $E_P$ does not
depend on $\ki$. Hence in the sphere or in the hyperbolic plane
this relation  has the same form as in Euclidean space; something
similar happens in the Kepler problem in Riemannian curved spaces
\cite{CRS05}. Third, both the method and the results obtained show
a close similarity with the Euclidean ones. The classical and well
known change of variable $r\, {\to}\,\uRegPar=1/r$ admits as a
generalization the $\ki$--dependent change $r\,{\to}\,\uRegPar =
1/\CKTr$ which affords a significant simplification for all values
of $\ki$; of course this change reduces to the Euclidean one
$r\,{\to}\,\uRegPar=1/r$ for $\ki\=0$.

\subsubsection{Method II: Equation of Binet }

The expression (\ref{ckho07:CKNoeMomCPol}) of the angular momentum
$\CKJ$ determines a relation between the differentials of the time
$t$ and the angle $\phi$
$$
   \CKJ\,dt = \CKSrSq\,d\phi \,.
$$
The corresponding relation between the derivatives with respect to
$t$ and $\phi$  is
$$
   \frac{d}{dt} = \Bigl(\frac{\CKJ}{\CKSrSq}\Bigr)\frac{d}{d\phi} \,,
$$
so that the second derivative with respect to $t$ is given by
$$
   \frac{d^2}{dt^2} = \Bigl(\frac{\CKJ}{\CKSrSq}\Bigr)\frac{d}{d\phi}
   \Bigl[\Bigl(\frac{\CKJ}{\CKSrSq}\Bigr)\frac{d}{d\phi}\Bigr] \,.
$$
Introducing this in the radial equation $\ddot{r} = \kii \CKSr \CKCr 
\dot{\phi}^2 - \frac{d\CKPot}{dr}$, it becomes
$$
   \Bigl(\frac{\CKJ}{\CKSrSq}\Bigr)\frac{d}{d\phi}
   \Bigl[\Bigl(\frac{\CKJ}{\CKSrSq}\Bigr)\frac{dr}{d\phi}\Bigr]
   = \frac{\CKJ^2}{\CKTr}\Big( \frac{1}{\CKSrSq} \Big) - \frac{d\CKPot}{dr}\,.
$$
This equation can be simplified in two steps: the
term in brackets in the l.h.s.\ can be rewritten by making use of
$$
\Bigl(\frac{\CKJ}{\CKSrSq}\Bigr)\frac{dr}{d\phi} =
-\CKJ \frac{d}{d\phi} \left( \frac{1}{\CKTr} \right)
$$
to obtain
$$
-\CKJ^2 \frac{d^2}{d\phi^2} \left( \frac{1}{\CKTr} \right) =
\kii \CKJ^2 \frac{1}{\CKTr} + \frac{d\CKPot}{dr}\,,
$$
and then we introduce the change $r\to \uRegPar$ with the
potential $\CKPot(\uRegPar)$ considered as a function of
$\uRegPar$. In this way we arrive at the differential equation of
the orbit
$$
\frac{d^2 \uRegPar}{d\phi^2}  + \kii \uRegPar = - \frac{1}{\CKJ^2} 
\frac{d\CKPot}{d\uRegPar}
$$
that permits us to obtain $\phi$ as a function of $\uRegPar$ for the
given potential considered as function of $\uRegPar$:
$$
\phi - \phi_0 = \pm \int \Bigl\{ c -\,\Bigl(\frac{2}{\CKJ ^2}\Bigr) 
\CKPot - \kii \uRegPar^2
   \Bigr\}^{-(1/2)} d \uRegPar  \,.
$$
(A $\pm$ sign is pertinent if both signs of the angular momentum
along a given orbit are considered; notice this equation coincides
with the one obtained in the previous subsection) Let us now
particularize for the harmonic oscillator $\CKPot =\smallonehalf\,
\omega_0^2\,(1/\uRegPar^2)$ in the CK space $\CKspace$: the
equation itself reduces to a nonlinear equation of Pinney--Ermakov
type
$$
   \frac{d^2\, \uRegPar}{d\phi^2}+ \kii \uRegPar = 
\frac{\omega_0^2}{\CKJ ^2}\,\frac{1}{\uRegPar ^3}
$$
whose general solution, further to the parameter $\CKJ$ already
present in the equation  depends on two independent integration
constants $D, \phi_0$ and has the form
\begin{equation}
    \uRegPar = \sqrt{\mathstrut D - G \CKcc(2(\phi-\phi_0))}, \qquad
    G = \sqrt{D^2-\frac{\omega_0^2}{\kii \CKJ^2}}
\label{ckho07:CKHOSolBinet}
\end{equation}
which coincides with the general orbit  obtained before. The
differential equation of the orbit for the variable $\uRegPar$,
usually known in the Euclidean case as {\it Binet's Equation}, is
essentially preserved by the $\ki, \kii$--deformation. Indeed, for
the three classical Riemannian spaces ($\kii\=1$) the new variable
$\uRegPar = 1/r $ deforms to $\uRegPar = 1/\CKTr$ but the equation
by itself remains invariant.

\subsubsection{Method III: Superintegrability in parallel coordinates}

In term of the parallel coordinates $(u, y)$  the Lagrangian which
represents the motion of a particle under an harmonic oscillator
potential $\CKPot$ in the CK space $\CKspace$ is:
\begin{equation}
   \CKLag(u, y, v_u, v_y) = \smallonehalf \left( \CKCySq {v_u}^2 + 
\kii{v_y}^2 \right)
          - \smallonehalf\, \omega_0^2\,\CKPot(u,y)      \,,\qquad
   \CKPot(u,y) = \frac{\CKTuSq}{\CKCySq} + \kii \CKTySq
\end{equation}
The expression of the potential, which correspond to the function
on $\CKspace$ given previously in polar coordinates,  displays its
separability in $(u, y)$ coordinates. The potential reduces to
$$
   \left.\CKPot\right|_{{\bf S}^2} =  \smallonehalf\, \omega_0^2
   \left( \frac{\tan^2 u}{\cos^2 y} + \tan^2 y \right) {\quad}
    \left.\CKPot \right|_{{\bf H}^2} = \smallonehalf\,  \omega_0^2
   \left( \frac{\tanh^2 u}{\cosh^2 y} + \tanh^2 y \right)
$$
in the two particular cases of the {\it standard} unit sphere
${\bf S}^2$ and Lobachevsky plane ${\bf H}^2$, and to
$$
    \left.\CKPot \right|_{{\bf E}^2}= V = \smallonehalf\, \omega_0^2
   \left( u^2 + y^2 \right) \equiv
   \smallonehalf\, \omega_0^2
   \left( x^2 + y^2 \right)
$$
in the Euclidean case (where we recall the equality $u\equiv x$ for $\ki\=0$).

Since we have already solved the dynamics in polar coordinates
$(r,\phi)$, we can make use of the expressions relating parallel
with polar coordinates (where the positive `$u$ axis' $y=0$ is
taken to coincide with the polar axis $\phi=0$; see Appendix).
Then the orbit equation (taking $\phi_0=0$ to avoid inessential
complications)
$$
   \Bigl(E_P -\,\sqrt{\,E_P^2 - \kii\omega_0^2\,\CKJ^2\,}\Bigr)
   \,\CKTrSq\CKCphiSq \,+\,
   \Bigl(E_P +\,\sqrt{\,E_P^2 - \kii\omega_0^2\, \CKJ ^2\,}\Bigr)
   \,\CKTrSq\CKCphiSq  = \kii \CKJ ^2  \,,
$$
becomes when written in coordinates $(u,y)$
$$
   \Bigl(E_P -\,\sqrt{\,E_P^2 - \kii\omega_0^2\,J^2\,}\Bigr)
   \,\CKTuSq \,+\,
   \Bigl(E_P +\,\sqrt{\,E_P^2 - \kii\omega_0^2\,J^2\,}\Bigr)
   \,\Bigl(\,\frac{\CKTy}{\CKCu}\,\Bigr)^2  = \kii \CKJ ^2\,,
$$
or
\begin{equation}
\frac{1}{A^2}\CKTuSq + \frac{1}{B^2}\frac{\CKTySq}{\CKCuSq} = 1\,.
\label{ckho07:CKHOOrbPCPar1}
\end{equation}
The coefficients $A, B$ are related to the values of the constants
of motion. By using the identity
$$
\frac{\CKTuSq}{\CKCySq} + \kii \CKTySq = \CKTuSq + \kii 
\frac{\CKTySq}{\CKCuSq}\,,
$$
the energy constant
\begin{equation}
E = \smallonehalf ({\CKP_1^2} + \kii {\CKP_2^2} +
          \ki\kii {\CKJ^2}) + \smallonehalf \omega_0^2 \bigg( 
\frac{\CKTuSq}{\CKCySq} + \kii \CKTySq \bigg)\,,
\label{ckho07:CKHOEneCPar1}
\end{equation}
can be rewritten as a linear combination of the constants related
to the superintegrability of the harmonic oscillator:
$$
I_E = \frac12 (I_{\CKP_1^2} + \kii I_{\CKP_2^2} +
          \ki\kii I_{\CKJ^2})
$$
where the quadratic constants of the motion specific to the harmonic 
oscillator in the space $\CKspace$ are:
\begin{equation}
\begin{tabular}{ll}
$I_{\CKJ^2}=\CKJ^2$\\[3pt]
$I_{\CKP_1^2}=\CKP_1^2 + \CKW_{11}(r,\phi)$,
                    \qquad & $\CKW_{11}(u,y) = \omega_0^2 \CKTuSq$ \\[3pt]
$I_{\CKP_2^2}=\CKP_2^2 + \CKW_{22}(r,\phi)$
                    \qquad & $\CKW_{22}(u,y) = \displaystyle\omega_0^2 
\frac{\CKTySq}{\CKCuSq} $\\[3pt]
$I_{\CKP_1\CKP_2}=\CKP_1\CKP_2 + \CKW_{12}(u,y)$
                    \qquad & $\CKW_{12}(r,\phi) 
=\displaystyle\omega_0^2 \frac{\CKTu\CKTy}{\CKCu} $.
\end{tabular}
\end{equation}
As in polar coordinates, all the functions $\CKW_{11}(u,y), 
\CKW_{22}(u,y), \CKW_{12}(u,y)$ are defined for any CK space, and 
they do not vanish identically in none of them. In the Euclidean 
case, the CK momenta $\CKP_1,\CKP_2$ and
the constants of motion reduce as they  should to:
\begin{equation}
\left.\CKP_1\right|_{{\bf E}^2}=v_u 
\quad
\left.\CKP_2\right|_{{\bf E}^2}=v_y 
\quad
\left. I_{\CKP_1^2} \right|_{{\bf E}^2} \!\!\!=
        v_u^2 + \omega_0^2\, u^2,
\quad
\left. I_{\CKP_1\CKP_2}\right|_{{\bf E}^2} \!=
        v_u v_y + \omega_0^2\, u y,
\quad
\left. I_{\CKP_2^2} \right|_{{\bf E}^2} \!\!\!=
        v_y^2 +  \omega_0^2\, y^2.
\end{equation}
where once more we recall that for $\ki\=0$, we have $u\equiv x$.
These constants are the elements of the Fradkin tensor \cite{Fr65,
JaH40}.

Back to the CK general case, and defining:
\begin{eqnarray}
2 E_1 := I_{\CKP_1^2}=\CKP_1^2 + \omega_0^2 \CKTuSq \\[2pt]
2 E_2 := I_{\CKP_2^2}=\CKP_2^2 + \omega_0^2 \frac{\CKTySq}{\CKCuSq}
\end{eqnarray}
the energy can be written as
$$
E = E_1 + \kii E_2 + \smallonehalf \ki\kii \CKJ^2
$$
where $E_1, E_2$ are the curved analogues to the `partial'
energies associated to the one-dimensional harmonic motions whose
linear superposition provides the most general 2d Euclidean
harmonic oscillator motion; remind that a curved configuration
space is not an affine space, thus strictly speaking there is no a
well defined way to superpose motions. The value of the remaining
constant of motion $I_{\CKP_1\CKP_2}$ is equal to zero for orbits
with $\phi_0=0$ (this choice amounts to diagonalize the Fradkin
tensor, and $2E_1, 2E_2$ are the eigenvalues). In this case, the
relation (\ref{ckho07:CKHOdetF}) gives
$$
4 E_1 E_2 = \omega_0^2 \CKJ^2\,,
$$
henceforth implying
\begin{equation}
(E_1 + \kii E_2)^2 - (-E_1 + \kii E_2)^2  = {4 \kii E_1 E_2} = {\kii 
\omega_0^2 \CKJ^2}\,.
\label{ckho07:CKHORelsE1E2}
\end{equation}
The quantity $E_P$ introduced before reduces simply to the `sum' of the
two $E_1, E_2$ contributions
$$
E_P = E_1 + \kii E_2
$$
and some simple algebra leads to neater expressions for the relations
among physical constants and conic coefficients:
\begin{equation}
D = \frac{E_1+\kii E_2}{\kii \CKJ^2},
\qquad
G = \frac{-E_1+\kii E_2}{\kii \CKJ^2}
\qquad
   {A^2} =\frac{2 E_1}{\omega_0^2}
   \qquad
   {B^2} =\frac{2 E_2}{\omega_0^2} .
\end{equation}

Direct observation in the expressions for the `superintegrability
constants' leads also to the following identity, holding for all
$\CKspace$:
\begin{equation}
   \CKF_{22} \CKTuSq
   - 2 \CKF_{12}\CKTu \Bigl(\,\frac{\CKTy}{\CKCu}\,\Bigr)
   + \CKF_{11}\Bigl(\,\frac{\CKTy}{\CKCu}\,\Bigr)^2
       = \kii \CKJ^2 \,.
\end{equation}
This property is proved by direct computation, and can be
interpreted as a relation which must hold between the coordinates
$(u,y)$ along a given motion, so this is precisely the orbit
equation, which can thus be obtained {\it directly} from the
superintegrable character. This is the most general orbit, but if
the coordinate axes are chosen so that the Fradkin tensor is
diagonal, then $I_{\CKP_1\CKP_2}=0$ and the equation coincides, of
course, with (\ref{ckho07:CKHOOrbPCPar1}). In the particular
$\ki\=0, \kii\=1$ Euclidean case we  obtain
$$
   \CKF_{22} x^2 - 2 \CKF_{12} x y + \CKF_{11}  y^2 =\, \CKJ^2
$$
that, as it is well known, represents an ellipse in the Euclidean
plane ${\bf E}^2$.

\subsection{The period of the harmonic oscillator}

The Euclidean oscillator is the classical example of an
isochronous system,  with the same period $T = {2\pi}/{\omega_0}$
for all orbits. A natural question is whether or not the `curved'
oscillator also inherits this property. We discuss this problem
for the closed orbits in Riemannian configuration spaces, where
$\kii\>0$. The strategy consists in choosing the orbits with
$\phi_0=0$, and computing the time spent by the particle between
the points $\phi=0$ and $\phi=\frac{\pi}{2\sqrt{\kii}}$, which
will always be reached along a closed orbit. By symmetry reasons
this time equals one fourth of the orbit period.

The angular velocity, taken from the angular momentum first integral:
$$
\dot{\phi} \equiv \frac{d\phi}{dt} = \frac{\CKJ}{\CKSrSq}\,,
$$
can be rewritten by successively using the identity
$$
\frac{1}{\CKSrSq} = \frac{1}{\CKTrSq}  + \ki\,,
$$
and the orbit equation
$$
\frac{1}{\CKTrSq} =
\frac{\omega_0^2}{2 E_1} \CKCphiSq +
\frac{\omega_0^2}{2 E_2} \CKSphiSq \,,
$$
as
$$
\frac{1}{\CKJ }\frac{d\phi}{dt} =
\frac{\omega_0^2}{2 E_1} \CKCphiSq +
\frac{\omega_0^2}{2 E_2} \CKSphiSq +
\ki \,,
$$
where by introducing double angles and simplifying we get:
$$
\frac{4 \kii E_1 E_2}{\CKJ \omega_0^2}\frac{d\phi}{dt} =
\kii \CKJ \frac{d\phi}{dt} =
\Big( E_1 + \kii E_2 + \ki \kii \CKJ^2 \Big) + \Big( -E_1 + \kii E_2 
\Big) \CKcc(2\phi)\,.
$$
By integrating along one fourth of a complete closed orbit:
$$
T = 4 \int_0^{\frac{\pi}{2\sqrt{\kii}}} dt = 4 \kii \CKJ 
\int_0^{\frac{\pi}{2\sqrt{\kii}}} \frac{d\phi}{\alpha + \beta \cos(2 
\sqrt{\kii} \phi)}\,,
$$
where $\alpha$ and $\beta$ denote the following expressions:
$$
\alpha:=\Big( E_1 + \kii E_2 + \ki \kii \CKJ^2 \Big)
\qquad
\beta:= \Big( -E_1 + \kii E_2  \Big) \,.
$$
The change $\zeta=2\sqrt{\kii} \phi$ makes the integration elementary
$$
T = 2 \sqrt{\kii} \CKJ \int_0^\pi \frac{d\zeta}{\alpha + \beta 
\cos\zeta} = 2 \sqrt{\kii} \CKJ \frac{\pi}{\sqrt{\alpha^2-\beta^2}} 
\,.
$$
Now some algebra and the use of the relations
(\ref{ckho07:CKHORelsE1E2}) leads to
$$
T = \frac{2\pi}{\omega_0} \frac{1}{\sqrt{1+\frac{2\ki E}{\omega_0^2}}} \,,
$$
an {\it exact} result whose expansion in powers of the curvature
has the correct Euclidean period as the zero-th order term, with a
fractional first-order correction by the dimensionless quotient
$\ki E/\omega_0^2$:
$$
T = \frac{2\pi}{\omega_0} \Big( 1 - \frac{E}{\omega_0^2}\ki + \cdots 
\Big), \qquad
\left. T \right|_{\ki\=0} = \frac{2\pi}{\omega_0} \,.
$$

Thus when the configuration space has curvature the period ceases
to be the same for all orbits (as it was to be expected), yet it
depends {\it only} on the total energy; this property is
characteristic for all closed orbits of superintegrable systems.
On the sphere, the period tends to infinity only when the orbit
approaches the equator, where $E\to\infty$, but on the hyperbolic
plane one might expect the period to diverge precisely when the
orbit character changes from a closed orbit to scattering open
orbit. From the previous section we know this happens for
$E=E_\infty$ and this  is precisely the value which makes
$1+\frac{2\ki E}{\omega_0^2}=0$, and hence $T = \infty$. For
energies above $E_\infty$ the derivation as provided is not
strictly applicable (because in these cases the orbit does not
reach any point with $\phi=\frac{\pi}{2\sqrt{\kii}}$; this is
reflected in the formally imaginary result got for $T$ in these
cases).


\section{Conics in spaces of constant curvature}

In this section we give a brief geometric description of conics in
the CK spaces, focusing on the three classical Riemannian
($\kii\!>\!0$) constant curvature spaces ${\bf S}^2_{\ki}, {\bf
E}^2, {\bf H}^2_{\ki}$ and emphasizing only those aspects relevant
in relation with oscillator motion in these spaces. This
description is intended to be self-contained, but as far as conics
in these spaces are concerned, it is  also complementary to
comments made in our previous paper devoted to Kepler problem
\cite{CRS05}.

We start by recalling the basics. Any $\CKspace$ is a homogeneous
symmetric space, and any homogeneous symmetric space has a
canonical connection which is always unique and well defined.
Restricting to the homogeneous spaces $\CKspace$, and due to the
quasi-simplicity of the involved Lie algebras $\mathfrak{so}_{\ki,
\kii}(3)$, there is always an invariant metric  $\pmb{\mathsf{g}}$
in $\CKspace$ which is non-degenerate whenever $\kii\!\neq\!0$.
The canonical connection is always compatible with the metric.
Generically, i.e., for $\kii\!\neq\!0$, we have more: the
canonical connection of $\CKspace$ as symmetric space  coincides
with the Levi-Civita (metric) connection associated to the metric.
In the degenerate case $\kii\=0$ however, there are many
connections compatible with the (degenerate) metric, and the
canonical connection of the homogeneous symmetric space is singled
out among the many connections compatible with the metric.

By {\it lines} will mean the {\it autoparallels} of this canonical
connection. Generically (when $\kii\!\neq\!0$), these  coincide
with the extremal curves of the length functional. In the
Lorentzian case there are two generic types of geodesics:
time-like, with real length and space-like, with `imaginary'
length relative to the main metric $\pmb{\mathsf{g}}$
(alternatively one may think of another space-like length along
space-like curves, corresponding to the `companion metric'
$\pmb{\mathsf{g}}/\kii$; the space-like curves have a real
`length'   as computed relative to this metric yet this choice
will not be used in this paper). Thus from now on, all distances
between points or between a point and a line will refer to lengths
along geodesics measured relatively to the main metric (hence when
$\kii\<0$ distances might be either real, vanishing or pure
imaginary).

The {\it geometric} definition of conics, which makes sense in any
2-d space of constant curvature $\ki$ and non-degenerate metric
($\kii\!\neq\!0$) involves {\it focal elements}, i.e., either
points which are assumed oriented, or lines which can be oriented
and co-oriented. In any such space, and by definition:

An {\it ellipse/hyperbola} will be the set of points with a constant
{sum/difference} $2a$ of distances $r_1, r_2$ to two fixed points $F_1,
F_2$ separated a distance $2f$ and called {\it foci}.

A {\it parabola} will be the set of points with a constant
{sum/difference} of distances $r_1, \tilde{r}_2$ to a fixed point
$F_1$, called {\it focus}, and to a fixed line $f_2$, called {\it
focal line}, ($\tilde{r}_2$ is assumed to be oriented); the {\it
oriented} distance between $F_1$ and $f_2$ plays here the role of
focal separation.

An {\it ultraellipse/ultrahyperbola} will be the set of points
with a constant {sum/difference} $2a$ of oriented distances
$\tilde{r}_1, \tilde{r}_2$ to two fixed intersecting lines $f_1,
f_2$ separated by an angle $2F$ and called {\it focal lines}.

In the generic CK 2-d space of constant curvature $\ki\!\neq\!0,
\kii\!\neq\!0$ these three pairs of curves, each pair sharing the
same focal elements, are the {\it generic} conics, and all the
remaining possible conics are either particular instances with
focal separation vanishing ($f=0, \f=0, F=0$) or limiting cases,
where some conic elements go to infinity (if possible at all);
both particular and limiting cases can be obtained as suitable
limits from the generic conics.

An important observation is that the Euclidean plane is {\it not
generic} neither among the complete family of CK spaces, nor among
the restricted family of Riemannian constant curvature planes.
Thus some Euclidean properties of conics  are very special and do
not provide a good viewpoint to assess the $\ki\!\neq\!0$
properties. For instance, in the {\it hyperbolic plane} ${\bf
H}^2_{\ki}$ a given ultraellipse, determined according to its
definition by a pair of intersecting focal lines, has not only
this pair of focal lines, but altogether  {\it three} such pairs,
the remaining two pairs being the non-intersecting pairs of focal
lines obtained by joining in the two possible ways the endpoints
of the (intersecting) initial pair. The distances between the two
members of each pair plays the role of focal separations in such
cases; and are in a suitable sense complementary (precisely, the
paralelism angles of half these distances are complementary in the
ordinary angular sense). A further important detail is that in
${\bf H}^2_{\ki}$, ultrahyperbolas for a given pair of
non-intersecting focal lines are ultraellipses for the matching
pair of non-intersecting focal lines (with the same end-points),
hence any ultraellipse can also be understood as an
ultrahyperbola, unlike ellipses and hyperbolas in ${\bf
H}^2_{\ki}$.

To make contact with the results in the physical part, we now draw
our attention precisely to the conics with centre, because the
orbit (\ref{ckho07:CKHOOrbitSPCPol}) of a particle in the
oscillator potential has a centre of symmetry at the origin of the
potential. In any CK space  ellipses/hyperbolas have centre, as do
ultraellipses/ultrahyperbolas as well as their common limits when
the `major' semiaxis go to infinity (if possible at all; this
might happen in the hyperbolic plane, where these limiting conics
are equidistants). Further analysis shows that conics appearing in
the oscillator problem are generically ellipses (but not
hyperbolas) and ultraellipses, with circles as the `equilateral'
case of an ellipse with equal semiaxes, and eventually
equidistants as the non-generic limiting orbits. Parabolas have
however no centre. This mean that oscillator orbits cannot be
parabolas in any CK space, and henceforth these conics will be
disregarded from now on.

In the sphere ${\bf S}^2_{\ki}$ there are no either points nor
lines at infinity, thus there are no limiting cases in the sense
they will appear in ${\bf H}^2_{\ki}$. This means that on the
sphere there is a single type of harmonic oscillator orbits:
generically ellipses on the sphere, with two limiting cases, lines
through the origin (for vanishing angular momentum) and circles
centred in the origin.

In the Euclidean plane ${\bf E}^2$ ultraellipses with any focal
angle $F$ are just pairs of parallel lines, in directions parallel
to the two bisectors of the focal angle; in the standard
description of Euclidean conics these are degenerate, and as far
as harmonic oscillator trajectories these are unphysical and
correspond to infinite energy and angular momentum.

The hyperbolic case ${\bf H}^2_{\ki}$ with {\it negative} constant
curvature is more interesting. The limiting conics relevant to the
oscillator problem are obtained when the major axis of the ellipse
tend to infinity, while the minor axis remains finite; the ellipse
tends to a limiting conic, which turns out to be a equidistant
curve to the major axis; this equidistant is also the limiting
conic obtained from an ultraellipse when its `major axis' tend to
infinity. In ${\bf E}^2$ these limiting conics collapse precisely
to pairs of parallel lines, but here they appear for finite values
for $E$ and $\CKJ$.

\subsection{Analysis of the orbits}

\subsubsection{The general CK configuration space}

In the physical part (Section 3) we obtained, in any $\CKspace$,
the equation of the orbit (\ref{ckho07:CKHOOrbitSPCPol}), which
contains three free parameters. One of them, $\phi_0$, can be set
to zero by taking appropriately the direction for the origin of
angles. Thus the orbit equation is:
\begin{equation}
   \frac{1}{\CKTrSq} = \frac{\CKccSq(\phi)}{A^2} + \frac{\CKssSq(\phi)}{B^2}\,,
\label{ckho07:CKHOOrbPCPol}
\end{equation}
which depends on two constants $A$ and $B$. In the Euclidean case,
this equation reduces to:
\begin{equation}
   \frac{1}{r^2} = \frac{\cos^2(\phi)}{A^2} + \frac{\sin^2(\phi)}{B^2}\,,
   \label{ckho07:EucHOOrbPCPol}
\end{equation}
which is an ellipse with semiaxes $A, B$. Again in the general CK 
space, the `physical' constants $E, E_1, E_2, \CKJ$ are related with 
the parameters $A, B$ by the relations
\begin{equation}
   {A^2} =\frac{2 E_1}{\omega_0^2}\,,
   \qquad\qquad
   {B^2} =\frac{2 E_2}{\omega_0^2}\,,
   \qquad\qquad
   A^2 B^2 = \frac{\CKJ^2}{\omega_0^2}.
\end{equation}
The type of the orbit as a conic in $\CKspace$ depends on the
space itself (the values of $\ki, \kii$) and on the constants $E,
E_1, E_2, \CKJ$, but the explicit form of the relations among $E,
E_1, E_2, \CKJ$ and $A, B$ turns out to be independent of the CK
parameters $\ki, \kii$; for the Euclidean case $\ki\=0, \kii\=1$
they reduce to the well known expressions, and in this case the
coefficients $A, B$ in (\ref{ckho07:EucHOOrbPCPol}) are directly
the orbit semiaxes. The total energy and `partial' translational
energy are given by
\begin{equation}
E = E_1 + \kii E_2 + \smallonehalf\ki\kii \CKJ^2,
\qquad
E_P = E_1 + \kii E_2 .
\label{ckho07:CKHOFVersusJE}
\end{equation}
The last physical constant is not the total energy, though it
coincides with the total energy in all flat configuration space);
it can be considered as a kind of `translational' part of the
energy.

\subsubsection{Orbits in a Riemannian CK configuration space}

We now discuss in more detail the standard Riemannian case
$(\kii\=1)$, with special emphasis in the two nonzero curvature
cases. The type of the orbit depends on the values of the
constants $A, B$. We consider altogether the family of all conics
whose equation, using  polar coordinates in ${\bf S}^2_\k, {\bf
E}^2, {\bf H}^2_\k$, with the focal symmetry axis of the conic
orbit as $\phi=0$, may be written in the form:
\begin{equation}
\frac{1}{\CKTrSq} =
      \frac{\cos^2(\phi)}{A^{2}} + \frac{\sin^2(\phi)}{B^{2}}\,,
\label{ckho07:RieHOOrbPCPol}
\end{equation}
with $A>B>0$. When $\ki\>0$ this equation describes spherical
ellipses, with spherical circles as the particular case $A=B$ and
with a spherical line (the equator) when $A=B=\infty$. When
$\ki\<0$, and depending on the values of $A, B$, this curve which
is always a conic might be generically either an ellipse or an
ultraellipse in ${\bf H}^2_{\ki}$. Before discussing each case
separately, notice that in the three Riemannian cases the
dependence on the polar angle $\phi $ is exactly the same as in
the Euclidean case. The quantities $A, B$ are related to the
lengths of the two semiaxes, but as one could expect, the details
depend on the sign of $\ki$: if $\ki\>0$ the range of values of
$\CKTr$ is the whole real line (completed with $\infty$), but when
$\ki\<0$, the values of $\CKTr$ are confined to the interval
$[-1/\sqrt{-\ki},1/\sqrt{-\ki}]$, which reduces to
$[0,1/\sqrt{-\ki}]$ for positive values for $r$. The values for
$r$ at $\phi=0$ and $\phi=\pi/2$ will play the role of major $a$
and minor $b$ semiaxis of the conic; then the relation between the
constants $A, B$ and the geometric semiaxes is
\begin{equation}
\CKTa=A, \qquad \CKTb=B\,.
\label{ckho07:CKHORelsABab}
\end{equation}
These relations identify the geometric meaning of the constants
$A, B$. In the non-negative curvature case, $\ki\geq0$, any $A, B$
will determine uniquely $a, b$, because the range of the circular
and parabolic tangent is the whole real line. But this is not so
when $\ki\<0$, for then this equation will define real values of
$a, b$ only when $A, B < 1/\sqrt{-\ki}$. If both $A, B$ are larger
than $1/\sqrt{-\ki}$, then the curve determined by the equation
(\ref{ckho07:RieHOOrbPCPol}) is empty (as the values in the r.h.s
are never got for any real value of $r$). Hence, we shall always
assume $B < 1/\sqrt{-\ki}$, so the only alternatives to study in
the hyperbolic plane ${\bf H^2}$, according to  $A$ is smaller
than, equal to or larger than $1/\sqrt{-\ki}$.

We remark that in the Riemannian case the origin of angles can
always be chosen so that $A>B$, and the major axis is precisely
the focal axis. We shall assume this choice; remark however that
in the Lorentzian case $\kii\<0$ there would be two cases to be
discussed, in agreement with the existence of two kinds of
separation.

\medskip

\noindent$\bullet$  (Standard) Spherical space $\ki\>0; \kii\=1$.
The polar equation of the orbit is:
\begin{equation}
   \frac{\ki}{\tan^2(\sqrt{\ki}r)} = \frac{\cos^2\phi}{A^2} + 
\frac{\sin^2\phi}{B^2}=\frac{\cos^2\phi}{\CKtSq{(a)}} + 
\frac{\sin^2\phi}{\CKtSq{(b)}}\,.
   \label{ckho07:EsfHOOrbCPol}
\end{equation}
For any values for $A, B$ with $A>B$, the values $a, b$ (with
$a>b$) are uniquely determined, and belong to the interval $[0,
\pi/(2\sqrt{\ki})]$. This curve is always closed, and is a
spherical ellipse with centre at the potential origin. If this
point is taken as the `pole', the complete orbit is contained in
one of the half-spheres bounded by the `equator', because the
r.h.s.\  of eq (\ref{ckho07:EsfHOOrbCPol}) is bounded by below, so
the value $r=\pi/(2 \sqrt{\ki})$ (for which $\frac{1}{\CKTrSq}=0$)
cannot be attained. For a fixed value of $\CKJ$ the minimal value
of the total energy corresponds to circles whose radius $r_m$
satisfies $\CKt(r_m) = \CKJ/\omega_0$ and have total energy
$E_{cir}=E(\CKJ):=\omega_0 \CKJ  + \smallonehalf \ki \CKJ^2$. The
values of the  energies of the possible motions for a given $\CKJ$
lie in the interval $[E(\CKJ),\infty]$; with our choice for origin
of angles $E_1 > E_2$. Any oscillator orbit through any point in
the upper/lower half-sphere centred at the origin is always
completely contained in that half-sphere, and from that viewpoint,
there is a single limiting orbit, the `equator' which corresponds
to $r_m =\pi/(2 \sqrt{\ki})$ and $a =b =\pi/(2 \sqrt{\ki})$ and $A
=B =\infty$; the physical constants for this orbit are $J=\infty,
E=\infty$.

\medskip

\noindent$\bullet$ (Standard) Hyperbolic space $\ki\<0; \kii\=1$.
Let us now consider the general case of arbitrary negative curvature. The
orbit equation is:
\begin{equation}
   \frac{-\ki}{\tanh^2(\sqrt{-\ki}r)} = \frac{\cos^2\phi}{A^2} + 
\frac{\sin^2\phi}{B^2}.
   \label{ckho07:HypHOOrbCPol}
\end{equation}

This curve is always a conic in the hyperbolic plane, but its type
depends on the values of $A, B$. This was to be expected from the
discussion for the effective potential, but it is worthy to look
at this situation from a more geometrical perspective.

Whitin the family of conics (\ref{ckho07:HypHOOrbCPol}) one can
expect ellipses with semiaxes $a, b$. These quantities should be
related with $A, B$ by means of (\ref{ckho07:CKHORelsABab}).  This
is only possible when $A^2<1/-\ki, B^2<1/-\ki$, for only if this
condition holds the values $A, B$ can be written as hyperbolic
tangents of actual values $a, b$, ranging in the interval $[0,
\infty]$:
$$
\CKtSq(a) = A^2, \qquad \CKtSq(b) = B^2.
$$
Its equation in polar coordinates is:
\begin{equation}
   \frac{-\ki}{\tanh^2(\sqrt{-\ki}r)} =
   \frac{\cos^2\phi}{\CKTaSq} + \frac{\sin^2\phi}{\CKtSq(b)}.
\label{ckho07:HypHOEllOrbCPol}
\end{equation}

What happens when $A^2$ or $B^2$ lie out of the former range? If
both $A^2>1/(-\!\ki),\ B^2>1/(-\!\ki)$, then for any $\phi$ it
would follow $\tanh^2(\sqrt{-\ki}r)>1$ a condition which cannot be
satisfied in any proper point of ${\bf H}^2_{\ki}$  and this
situation cannot produce any oscillator orbit. Then we are left
with the case $A^2>1/(-\!\ki),\ B^2<1/(-\!\ki)$. In this case the
conic (\ref{ckho07:HypHOOrbCPol}) is {\it not} an hyperbolic
ellipse, but a {\it ultraellipse}, and formally its semiaxis is
not an actual distance. To cater for these cases we introduce
another real quantity $\tilde{a}$, formally complementary to the
would-be semiaxis $a$, and which is related to $A$ by
$$
\frac{1}{-\ki\CKTaTl} =  A\,.
$$
Notice that when $\tilde{a}\in[0, \infty]$, the function
$1/({\ki^2\CKTaTlSq})$ ranges in the interval $[1/-\ki, \infty]$
and therefore, when considered altogether with $\CKTaSq$, the
union of the ranges of the two functions $\CKTaSq$ and
$1/(\ki^2\CKTaTlSq)$ fills the real line. In the following we will
refer to $\tilde{a}$ as the ultraellipse `semiaxis'; the polar
equation of the ultraellipse with semiaxes $\tilde{a}$ and $b$ is:
\begin{equation}
   \frac{-\ki}{\tanh^2(\sqrt{-\ki}r)} =
   \frac{\cos^2\phi}{\frac{1}{\ki^2\CKTaTlSq}} + \frac{\sin^2\phi}{\CKTbSq}
   =
   \ki^2\CKTaTlSq {\cos^2\phi} + \frac{\sin^2\phi}{\CKTbSq}\,.
\label{ckho07:HypHOUEllOrbCPol}
\end{equation}

This type of conic in ${\bf H}^2_{\ki}$ has no generic analogue in
the Euclidean plane; the Euclidean `limit' $\ki\=0$ of
(\ref{ckho07:HypHOUEllOrbCPol}) is a straight line, which would
correspond to an Euclidean oscillator orbit with infinite energy
and angular momentum. On the hyperbolic plane these oscillator
orbits reach the spatial infinity with finite values for energy
and angular momentum.

Therefore, in ${\bf H}^2_{\ki}$, the equation of the complete
family of conics we are considering is given, generically,  by one
of the two mutually exclusive possibilities:
\begin{equation}
      \frac{-\ki}{\tanh^2(\sqrt{-\ki}r)} =
      \frac{\cos^2\phi}{\CKTaSq} + \frac{\sin^2\phi}{\CKtSq(b)}, \qquad
      \frac{-\ki}{\tanh^2(\sqrt{-\ki}r)} =
      \k^{2}\CKtSq(\tilde{a}) {\cos ^2\phi} + \frac{\sin ^2\phi}{\CKtSq(b)}\,,
\label{ckho07:HypHOGenOrbCPol}
\end{equation}
where $b\in[0 \leq b <\infty]$ and either $0\leq a <\infty$ or
$\infty>\tilde{a}\geq0$. Both types of orbits have a common
limiting case, either $a\to\infty$ or $\tilde{a}\to\infty$, which
corresponds to the conic
\begin{equation}
\frac{-\ki}{\tanh^2(\sqrt{-\ki}r)}  =
\frac{\cos^2\phi}{\frac{1}{-\ki}} + \frac{\sin^2\phi}{\CKTbSq} =
      (-\ki)^2{\cos^2\phi} + \frac{\sin^2\phi}{\CKtSq(b)}\,.
\label{ckho07:HypHOSingOrbCPol}
\end{equation}
This conic is neither an ellipse nor an ultraellipse in the
hyperbolic plane, but can be obtained from either of these types
as the limit $a\to\infty$ or $\tilde{a}\to\infty$ while $b$ is
fixed. In the Euclidean case, the limiting ellipse is clearly a
straight line, equidistant to the major axis and with $b$ as
equidistance. In the hyperbolic plane, this conic is an
equidistant curve, with equidistance $b$ from the $x$-axis. This
means that in the case of the hyperbolic plane ${\bf H}^2_{\ki}$
the family of conics we are considering includes conics
intersecting the major axis at proper points (the ellipse), at
only at infinity (the equidistant) or not intersecting it at all
(the ultraellipse).

Of course the two expressions (\ref{ckho07:HypHOGenOrbCPol}) can
be used as well when $\ki\>0$ but then each of the two
alternatives covers actually all cases. And further, only the
first possibility has a sensible Euclidean limit because the
$\ki\to0$ limit of the $\tilde{a}$ family lies completely at the
infinity; the existence of this  family as a different set of
conics is specific to the hyperbolic plane.

In Section 3 we discussed the appearance of some energy and
angular momentum standards $E_\infty, \CKJ_\infty$ in the
hyperbolic plane. As it might be expected, these correspond to the
transition among different types of oscillator orbits. To be
precise, very  simple calculations show that in ${\bf H}^2$:

\noindent$\bullet$ All elliptic orbits have energies $E<E_\infty$
and angular momentum $\CKJ<\CKJ_\infty$

\noindent$\bullet$ The equidistant orbits have energy $E=E_\infty$
and angular momentum $\CKJ<\CKJ_\infty$

\noindent$\bullet$ The ultraelliptic orbits have energies
$E>E_\infty$ and the angular momentum $\CKJ$ can have any value.
The ultraellipses with $\CKJ<\CKJ_\infty$ are those with
$\CKt(b)<\CKTaTl$, while those with $\CKt(b)>\CKTaTl$ have
$\CKJ>\CKJ_\infty$. Separating these behaviours are the
`equilateral' ultraellipses with $\CKt(b)=\CKTaTl$, all of which
have $\CKJ=\CKJ_\infty$.

\noindent$\bullet$ The `largest circular orbit', with radius
$r=\infty$ is a common limit from elliptic orbits when
$a\to\infty, b\to\infty$ or from equidistant orbits, when $b\to
\infty$. This orbit  has energy $E=E_\infty$ and angular momentum
$\CKJ=\CKJ_\infty$


It is clear that when the total energy $E$ is smaller than
$E_{\infty}$, the trajectory (an ellipse) is bounded and the
motion is periodic, while for $E$  equal to or larger than
$E_{\infty}$, the motion is not periodic and the orbit goes to
spatial infinity.

\medskip

We finally give the relation between the two `geometric'
quantities, the ellipse (or ultraellipse) semiaxes $a$ (or
$\tilde{a}$) and $b$, the parameters $A, B$ appearing in the
canonical form of the orbit (\ref{ckho07:CKHOOrbPCPol}) and the
angular momentum $\CKJ$ and energies $E, E_1, E_2$:
\begin{equation} \omega_0^2 \CKtSq(a) =  \omega_0^2 A^2 \equiv
2E_1, \quad\hbox{ or }\quad \omega_0^2 \frac{1}{\ki^2
\CKtSq(\tilde{a})} =  \omega_0^2 A^2 \equiv 2E_1,
\end{equation}
\begin{equation}
\omega_0^2 \CKtttSq(b)=  \omega_0^2 B^2 \equiv 2E_2, \qquad
\end{equation}
\begin{equation}
\CKJ = \omega_0 \CKt(a) \CKttt(b),
\quad\hbox{ or }\quad
\CKJ = \omega_0 \frac{1}{-\ki \CKt(\tilde{a})} \CKttt(b),
\end{equation}
\begin{equation}
E = \smallonehalf \omega_0^2
\left\{ \frac{\CKtSq(a)}{\CKcSq(b)} + \kii \CKtttSq(b) \right\}
\quad\hbox{ or }\quad
E = \smallonehalf \omega_0^2
\left\{ \frac{1}{\ki^2 \CKtSq(\tilde{a})\CKcSq(b)}  + \kii 
\CKtttSq(b) \right\}.
\end{equation}


\section{Final comments and outlook}

We have discussed and completely solved the harmonic oscillator
problem simultaneously on the nine  2d spaces of constant
curvature and metric of either signature type. As stated in the
introduction, one of the fundamental characteristics of this
approach is the use of two parameters $\ki, \kii$ in such a way
that all properties we have obtained reduce to the appropriate
property for the system on each particular space. Generically
these spaces include either the sphere ${\bf S}^2$, the hyperbolic
plane ${\bf H}^2$, the De Sitter ${\bf dS}^\unomasuno$ or the
AntiDeSitter spaces ${\bf AdS}^\unomasuno$ when the corresponding
values of $\ki, \kii$ are appropriately set. Euclidean or
Minkowskian spaces arise as the very particular (but important)
flat cases $\ki\=0$ with a non-degenerate metric $\kii\!\neq\!0$.
So, we can sum up the results pointing out some important facts:

\begin{itemize}
\item  The harmonic oscillator is not a specific or special system
living only on the Euclidean space.  In any of the nine CK spaces
$\CKspace$ there is a  system with all the outstanding properties
of the Euclidean harmonic oscillator, worth of the name `curved
harmonic oscillator'.

\item  Motion in the `curved harmonic oscillator' potential on any
CK space is superintegrable.
\end{itemize}

The orbits of motion in this potential are always conics with
centre at the origin of potential, in all the CK spaces and in the
sense of the intrinsic geometry in $\CKspace$. These conics always
include ellipses, with straight lines trough the origin as
limiting cases for $\CKJ=0$, but other types of conics might also
appear in some spaces. For the three Riemannian spaces of constant
curvature, the situation is as follows:

\begin{itemize}
\item  In the sphere case, there is a single type of oscillator
orbits; all such orbits are ellipses and are confined to an
hemisphere centred in the oscillator centre; a `limiting' case is
the `equator' orbit, which appears as an ellipse with major axis
equal to half the equator's length (on the sphere this condition
implies both axis to be equal); this orbit has infinite energy and
angular momentum.

\item  In the Euclidean case (the $\ki\to0$ limit of the sphere
case), all oscillator orbits are ellipses, and a limiting case
happens when the orbit tends to a straight line not through the
origin, with infinite angular momentum and energy.

\item  In the hyperbolic plane, orbits with values of $E$ smaller
than $E_\infty$ are ellipses, and for them $\CKJ$ should be
smaller than  $\CKJ_\infty$. This ellipse reaches the spatial
infinity when $E= E_\infty$; this happens when the focal distance
goes to infinity but the minor axis stays finite, and the limiting
curve is an equidistant, with base curve the major axis and
equidistance distance equal to the minor semiaxis. For values of
$E>E_\infty$  the orbit is (a branch of) an ultraellipse, a conic
reaching the spatial infinity, and in this case all values of
$\CKJ$ (even larger than $\CKJ_\infty$) are allowed; in the limit
where $E, \CKJ$ go to infinity, the orbit tends to a hyperbolic
straight line, which appears here as a particular limiting case of
an ultraellipse.
\end{itemize}

Then, as compared with the Euclidean case, the new trait appearing
in the hyperbolic plane ${\bf H}^2_{\ki}$ is the `splitting' of
the Euclidean `last' singular straight line in a full family of
orbits: a `first' orbit reaching the spatial infinity, then a full
family of ultraelliptic orbits and finally a straight line orbit
(Figure 4). And the new trait in the sphere case is that the
Euclidean `last' singular orbit becomes the sphere equator, which
is also a straight line in the spherical geometry. Qualitatively,
this reminds the results in the Kepler problem, where the single
Euclidean parabolic orbit `splits` for $\ki\<0$ into a full family
of Kepler orbits in ${\bf H}^{2}$, bordered by two different
limiting curves, an horoellipse as a limiting form of ellipses and
a horohyperbola as a limiting form of hyperbolas, with a full
interval of parabolas between them. The separating role in the
oscillator problem is played by equidistants.

Of course, since the nine CK manifolds $\CKspace$ are
geometrically very different, many dynamical properties display
some differences according to some distinguishing properties of
the manifolds; nevertheless, all these differences can be traced
back ultimately to characteristics related to the signs of the
basic parameters $\ki, \kii$ so there is a unique theory that is
simultaneously valid for all the cases, for any value of the
curvature and any signature type.

The analysis of the orbits in the curved harmonic oscillator
leads, in a natural way, to the theory of conics on spaces of
constant curvature and any signature type. Although Sects.\ 3 and
4 were mainly concerned with dynamical questions, Sect.\ 5 was
written emphasizing its geometrical character. It is clear that
the theory of conics on spaces of constant curvature is a
geometrical matter of some importance deserving a deeper study
within this CK formalism that we hope to present elsewhere; in the
three Riemannian spaces of constant curvature conics have been
discussed, at different depths and from differing viewpoints, in
papers dating from a century or more (see for instance
\cite{KleinBook}, p.\ 229 or \cite{CooBook}). Of course, none of
these papers discussed the case of Lorentzian spaces, where it
appears that this theory has been never presented systematically.


\section*{Appendix: Geodesic coordinates on two-dimensional manifolds}

Consider the generic CK space $\CKspace$, for any values of $\ki,
\kii$. When $\kii$ is positive it may be reduced to 1, and then
this family includes the three constant curvature 2d Riemannian
spaces $V^2_{\ki}$. When $\kii$ is negative, it may be reduced to
$-1$ and the spaces are Lorentzian manifolds of constant curvature
$L^2_{\ki}$. We now describe the two types of coordinates employed
in the paper. Choose any point $O$ a point on $\CKspace$ and let
$l_1$ be an oriented geodesic  (time-like if $\kii$ is
non-positive, generated by $P_1$) through $O$, and $l_2$ the
oriented geodesic orthogonal to $l_1$ through $O$ (hence
space-like if $\kii$ is non-positive and generated by $P_2$) (see
Figs. 5 and 6)

\subsection*{I: Geodesic polar coordinates}

For any point $Q$ in some suitable domain (with time-like
separation to $O$ in the lorentzian case), there is a unique
geodesic $l$ joining $Q$ with $O$. The (geodesic) polar
coordinates $(r,\phi)$ of $Q$, relative to the origin $O$ and the
positive geodesic ray of $l_1$, are the distance $r$ between $Q$
and $O$ measured along $l$, and the angle $\phi$ between $l$ and
the positive ray $l_1$, measured around $O$ (Figure 5). These
coordinates are defined in some domain not extending beyond the
cut locus of $O$, are singular at $O$; when $\kii\>0$, $\phi$ is
discontinuous on the positive ray of $l_1$, where there is a jump
of $2 \pi/\sqrt{\ki}$, but if $\kii\<0$ then the range of $\phi$
covers all real line, and there are no jumps, at the price of the
coordinates being singular on the isotropes through $O$. In the
sphere the domain of polar coordinates only fails to include two
points: $O$ and its antipodal; for the hyperbolic plane it covers
all the space except the point $O$; for the Anti De Sitter this
covers the domain with time-like separation to $O$, save $O$ and
its antipodal, etc.

The expression for the differential element of distance $dl$
is given by
$$
   ds^2 = d r^2 + \kii \CKSrSq \,d{\phi}^2 \,,
$$
so that in the standard Euclidean case ($\ki\=0, \kii\=1$) we get
$\left.ds^2\right|_{{\bf E}^2} = d r^2 + r^2\,d{\phi}^2$.

\subsection*{II: Geodesic parallel coordinates}

For any point $Q$ in some suitable domain, there is a unique
geodesic $l'_2$ through $Q$ and orthogonal to $l_1$, intersecting
$l_1$ at a point denoted $Q_1$; the geodesic $l'_2$ is space-like
in the Lorentzian case. Alternatively, for $Q$ in some suitable
domain, there is a unique geodesic $l'_1$ through $Q$ and
orthogonal to $l_2$ intersecting $l_2$ at a point denoted $Q_2$;
the geodesic $l'_1$ is time-like in the Lorentzian case. The
points $Q_1$, $Q_2$ can be considered as the `orthogonal
projections' of $Q$ on the lines $l_1, l_2$ (recall that
$\CKspace$ is in general not an affine space). Then we can
characterize the point $Q$ by (Figure 6)

\begin{enumerate}
\item Two coordinates $(u,y)$. The coordinate $u$ is the canonical
parameter of the element in the one-dimensional subgroup of
translations along $l_1$, generated by $P_1$ and with label $\ki$,
which brings $O$ to $Q_1$; this value  coincides with the distance
between $O$ and $Q_1$ computed along $l_1$ with the CK space
metric. The coordinate $y$ is the canonical parameter of the
element in the one-dimensional subgroup of translations along
$l'_2$, generated by $e^{uP_1}P_2e^{-uP_1}$ and with label
$\ki\kii$, which brings $Q_1$ to $Q$; this value is related with
the distance between $Q_1$ and $Q$ computed along $l'_1$ with the
CK space metric by a factor $\sqrt{\kii}$ (remark in the
Lorentzian case, $y$ is always real, yet both a space-like
separation and $\sqrt{\kii}$ are pure imaginary).

\item Two coordinates $(x,v)$. The coordinate $v$ is the canonical
parameter of the element in the one-dimensional subgroup of
translations along $l_2$, generated by $P_2$ and with label
$\ki\kii$, which brings $O$ to $Q_2$; this value is related with
the distance between $O$ and $Q_2$ computed along $l_2$ with the
CK space metric by a factor $\sqrt{\kii}$, so that $v$ is always
real even in the Lorentzian case. The coordinate $x$ is the
canonical parameter of the element in the one-dimensional subgroup
of translations along $l'_1$, generated by $e^{vP_2}P_1e^{-vP_2}$
and with label $\ki$, which brings $Q_2$ to $Q$; this value
coincides with the distance between $Q_2$ and $Q$ computed along
$l'_1$ with the CK space metric.
\end{enumerate}

In the first case we have the parallel coordinates of $Q$ relative
to $(O,l_1)$ and in the second case relative to $(O,l_2)$ (Figure
6). In the $(u,y)$ system the curves `$u={\rm constant}$' are
geodesics meeting ortoghonally the `base' geodesic $l_1$, and the
curves `$y={\rm constant}$' are equidistant lines to the base
$l_1$, and intersect orthogonally
  $u={\rm constant}$.
In the $(x,v)$ system the curves '$v={\rm constant}$' are
geodesics and the lines `$x={\rm constant}$' are equidistant to
$l_2$. Notice that in the general case, with non-zero curvature
$\ki\!\neq\!0$ we have $u{\ne}x$ and $v{\ne}y$; only in the case
of flat spaces do the equalities $x=u, v=y$ hold.

The $(u,y)$ and $(x,v)$ expressions for the differential element
of distance $ds^2$ are given by \cite{Klin78}
$$
   ds^2 = \CKCySq\,du^2 + \kii dy^2 \,,\quad\quad
   ds^2 = dx^2 + \kii \CKCxSq\,dv^2 \,,
$$
The three coordinate systems can be related by the general
formulae of trigonometry in the CK space $\CKspace$
\cite{HeOrSa00}.
$$
   \CKTu = \CKTr\CKCphi \,,{\quad}
   \CKSy = \CKSr\CKSphi \,,{\quad}
   \CKCu \CKCy = \CKCr \,.
$$

In particular, the two parallel coordinate systems coincide when
$\ki\=0$ reducing to $ds^2 = dx^2 + \kii dy^2$; in the standard
Euclidean case with $\kii\=1$ this reduces further to $ds^2 = dx^2
+ dy^2$. In this article we have made use of only the $(u,y)$
coordinates, but all the CK formulation can be easily expressed
also  in the $(x,v)$ coordinates.


\section*{\bf Acknowledgments.}
Support of projects E24/1 (DGA), MTM-2005-09183, MTM-2006-10531,
and VA-013C05 is acknowledged.



\def\otherrefs#1{}

\small





\vfill\eject

\section*{Figures and figure captions}

{\hfill\includegraphics[width=300pt]{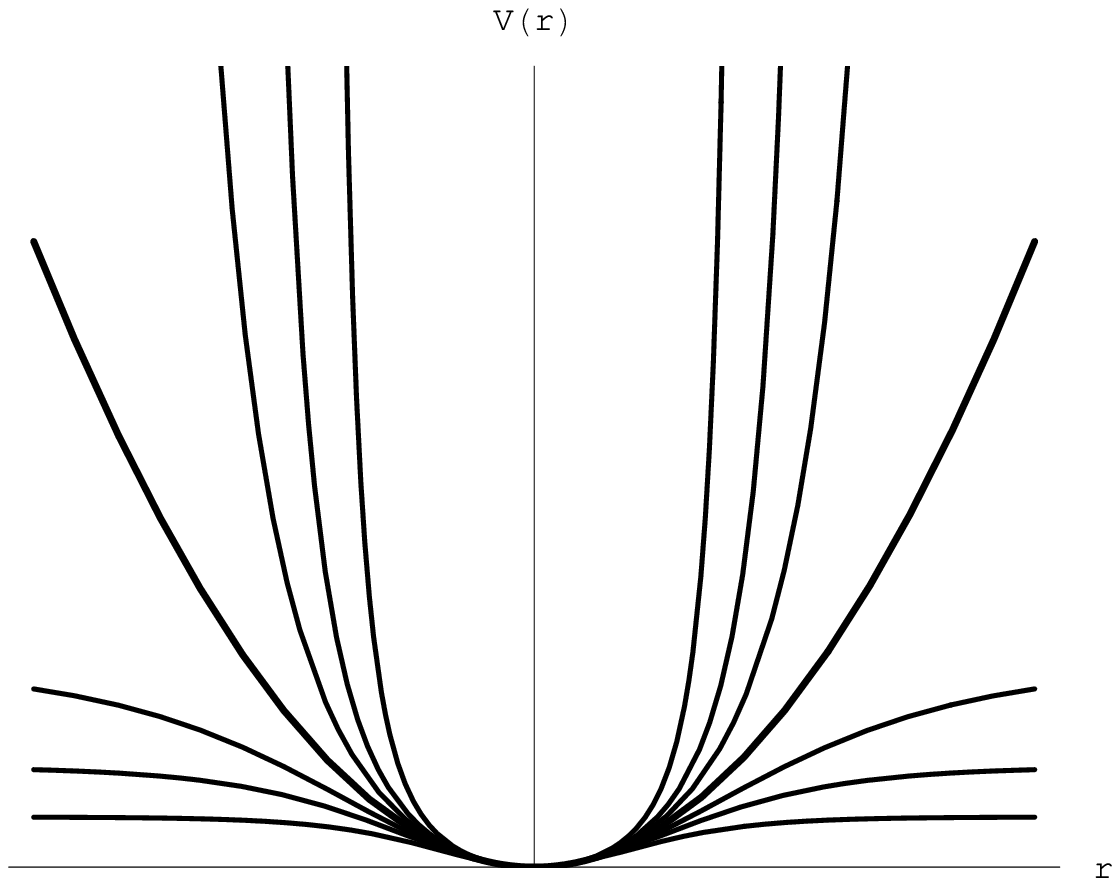}\hfill}

\noindent{\sc Figure 1}.{\enskip} Plot of the harmonic oscillator
Potential $\CKPot(r)$ as a function of $r$, for several values for
the curvature. Upper curves correspond to positive curvature
$\k=2, 1, 0.5$; the slightly thicker line to the Euclidean plane
$\k=0$; the lower curves to negative curvatures $\k=-0.5, -1, -2$.
All the functions have the same quadratic behaviour around $r=0$
and the quadratic Euclidean function appears in this formalism as
making a separation between two different behaviours (an infinite
wall at finite $r$ versus a finite plateau at infinite $r$). In
the case $\kii>0$ the natural range of the radial coordinate
includes only positive values, but when $\kii<0$ positive and
negative $r$ appear naturally.

\bigskip\bigskip\bigskip

{\hfill\includegraphics[width=200pt]{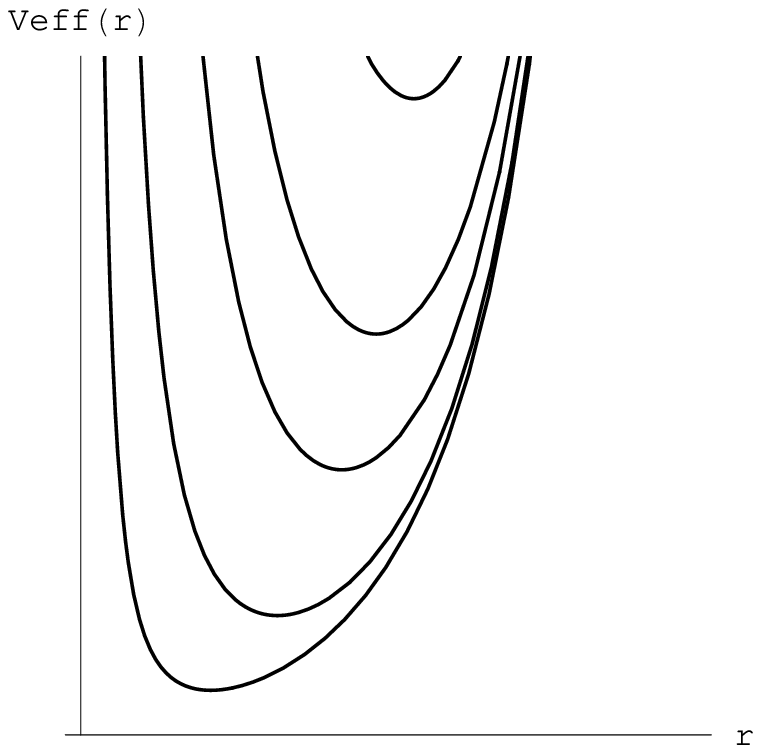}\hfill}

\noindent{\sc Figure 2}.{\enskip} Plot of the effective potential
$\CKPoteff(r)$ as a function of $r$ in the standard Riemannian
positive curvature case ($\ki\=1, \kii\=1$) depicted for several
values for $\CKJ$. All these potentials are asymmetric wells with
two infinte walls at $r=0$ and $r=\pi/2$.

\bigskip\bigskip\bigskip

{\hfill\includegraphics[width=250pt]{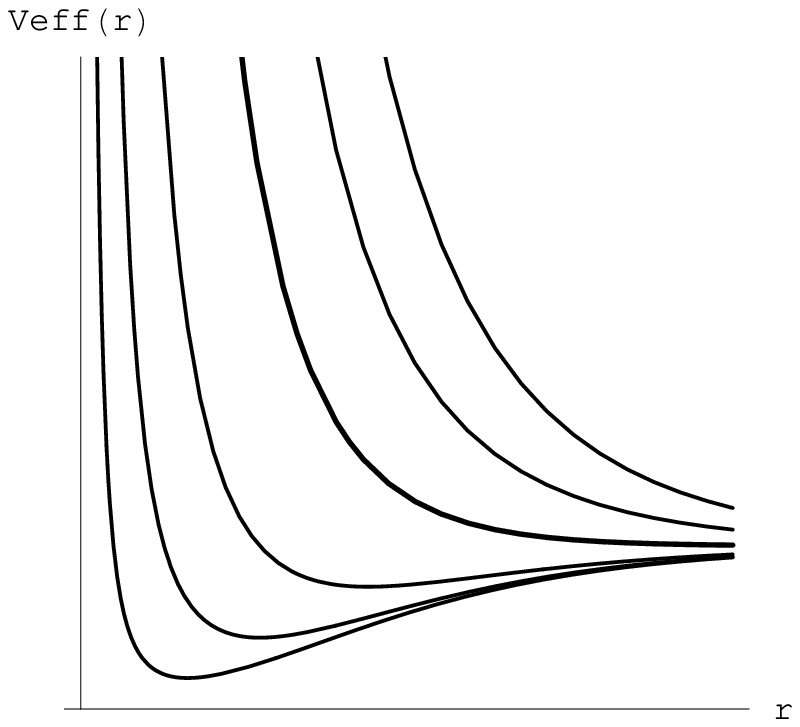}\hfill}

\noindent{\sc Figure 3}.{\enskip} Plot of the effective potential
$\CKPoteff(r)$ as a function of $r$ in the standard Riemannian
negative curvature case ($\ki\=-1, \kii\=1$) depicted for several
values for $\CKJ$. The angular momentum standard $\CKJ_\infty$
corresponds to the slightly thicker curve, where behaviour of the
effective potential changes. Curves for values of $\CKJ$ greater
(resp.\ lower) than $\CKJ_\infty$ appear above (resp.\ below) this
curve and correspond to an equivalent potential without (resp.\
with) a minimum.

\bigskip\bigskip\bigskip

\noindent
\def\figIVsize{140pt}
\includegraphics[width= \figIVsize]{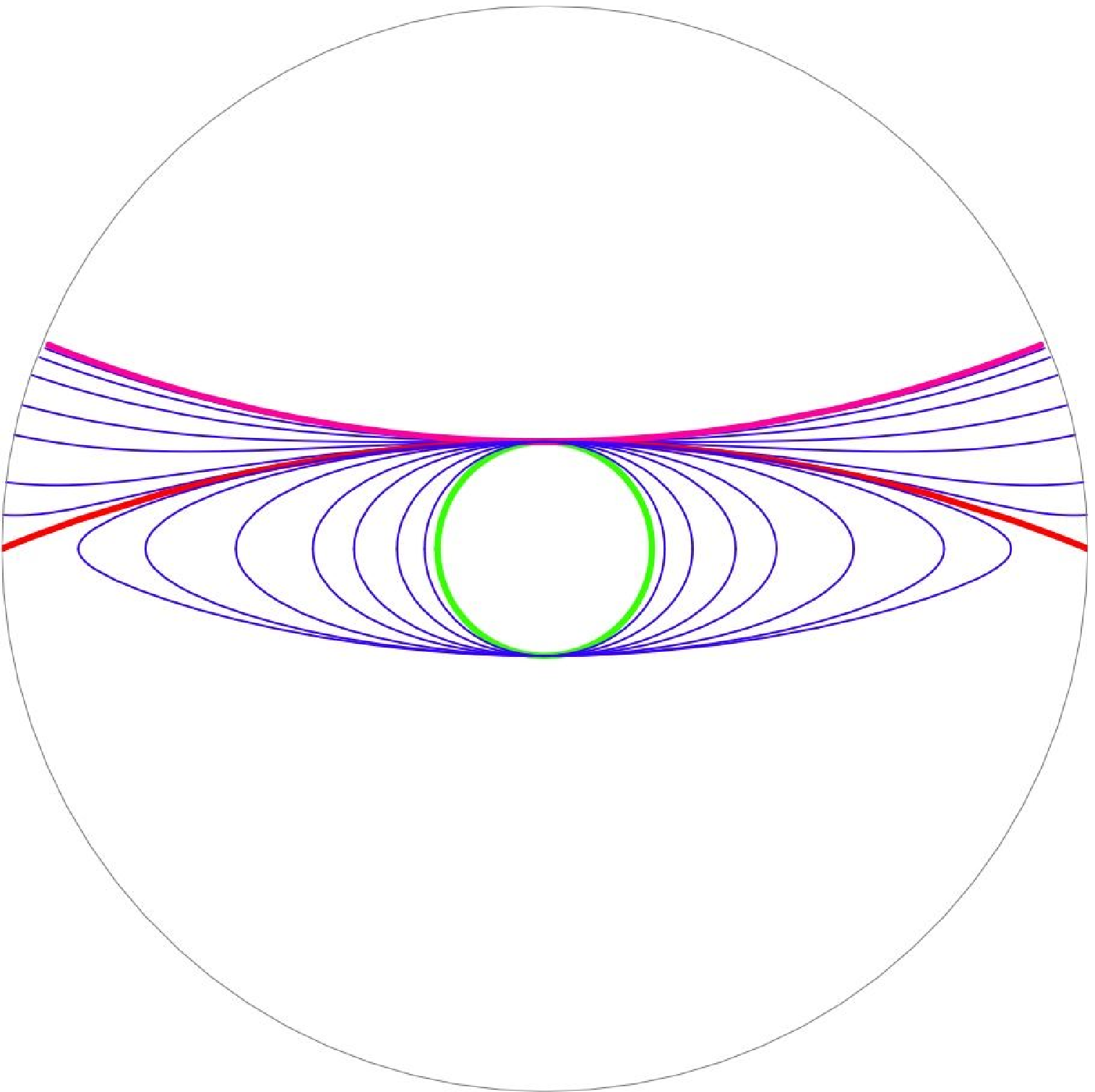}
\includegraphics[width= \figIVsize]{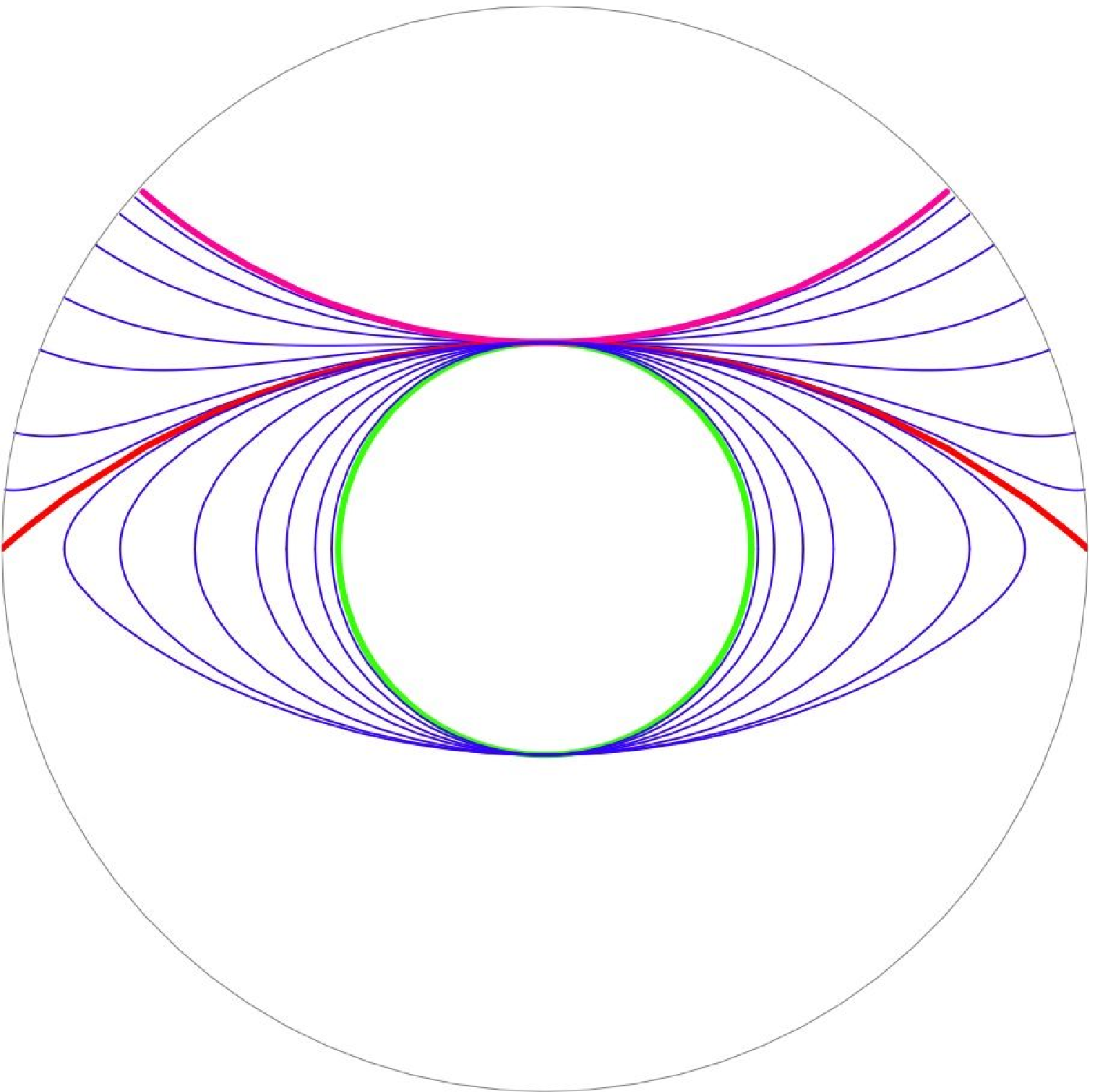}
\includegraphics[width= \figIVsize]{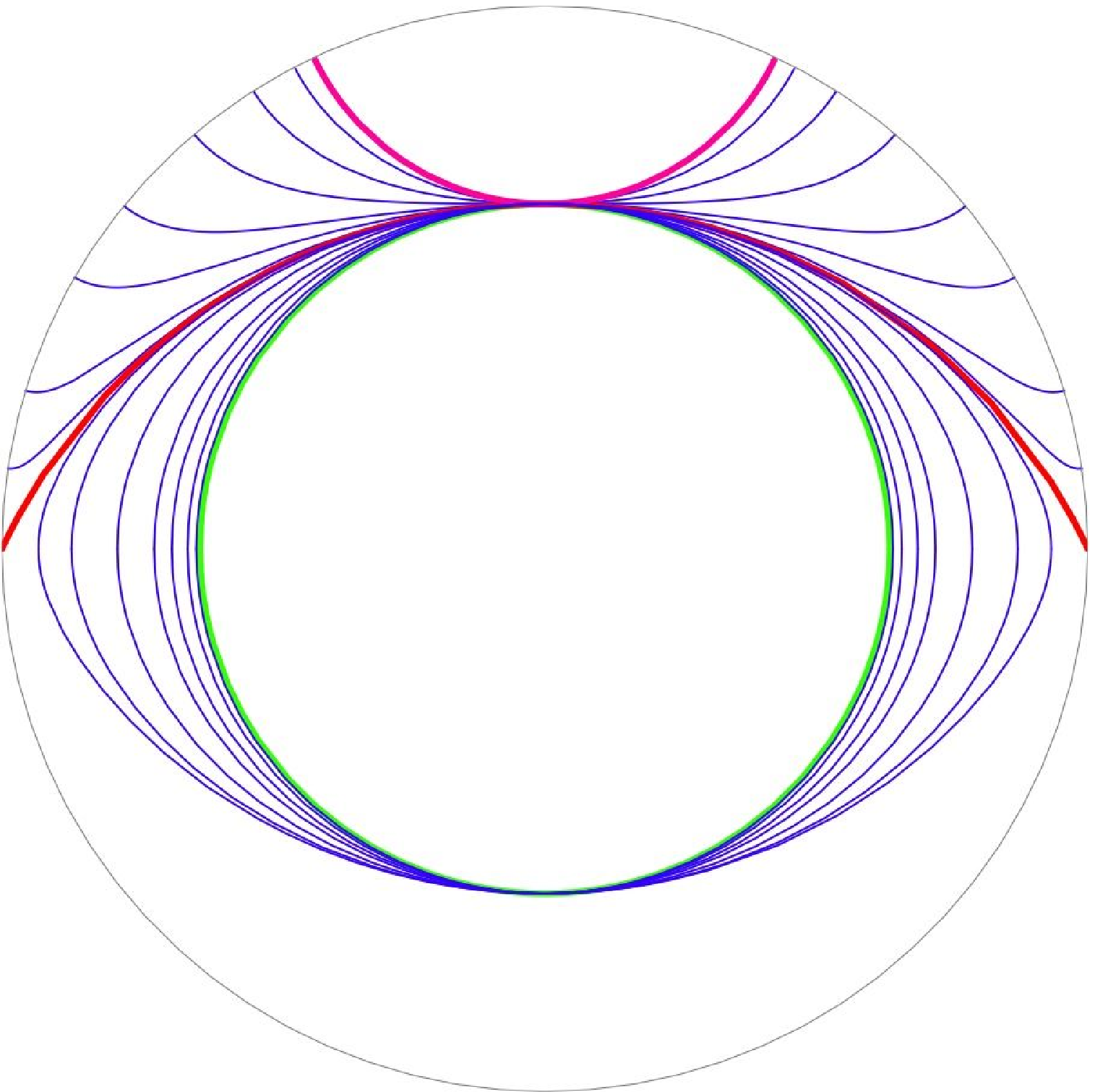}

\noindent{\sc Figure 4 abc}. {\enskip} Harmonic oscillator orbits
in a hyperbolic plane configuration space, depicted in the
conformal Poincare disk model. Each figure displays orbits with a
fixed value for the minor semiaxis $b$ (or equivalently, fixed
`partial energy' $E_2$) and several values for the major semiaxis
$a$, ranging from $a=b$ (circular orbit, in green), seven ellipses
for increasing values of $a$ (in blue), an equidistant curve for
$a=\infty$ or $\tilde{a}=\infty$ (in red), seven ultraellipses for
decreasing values of $\tilde{a}$ (in blue) and finally the
straight orbit for $\tilde{a}=0$ (in magenta). From left to right,
$b$ is ranging from `small' (Figure 4a)  to `large' (Figure 4c)
values. The potential centre is at the origin, which is a centre
of the conics. Colors have been chosen to represent particular and
limiting conics:  circle (green), equidistant (red) and straight
line (magenta). For ellipses the pair of focus (not marked) are on
the horizontal line; for the equidistant the foci are at infinity,
as well the focal lines, which are orthogonal to the horizontal
line $l_1$ at infinity; for the ultraellipses one set of focal
lines is orthogonal to the horizontal line. Notice only orbits
with total energy smaller than $E_\infty$ intersect the horizontal
line $l_1$ and come back to the initial point. Orbits with total
energy larger than this value are not closed and go to spatial
infinity. The two families in blue (ellipses and ultraellipses)
are the two generic behaviours, as explained in the text.

\bigskip\bigskip\bigskip
\newpage

{\hfill\includegraphics[width=400pt]{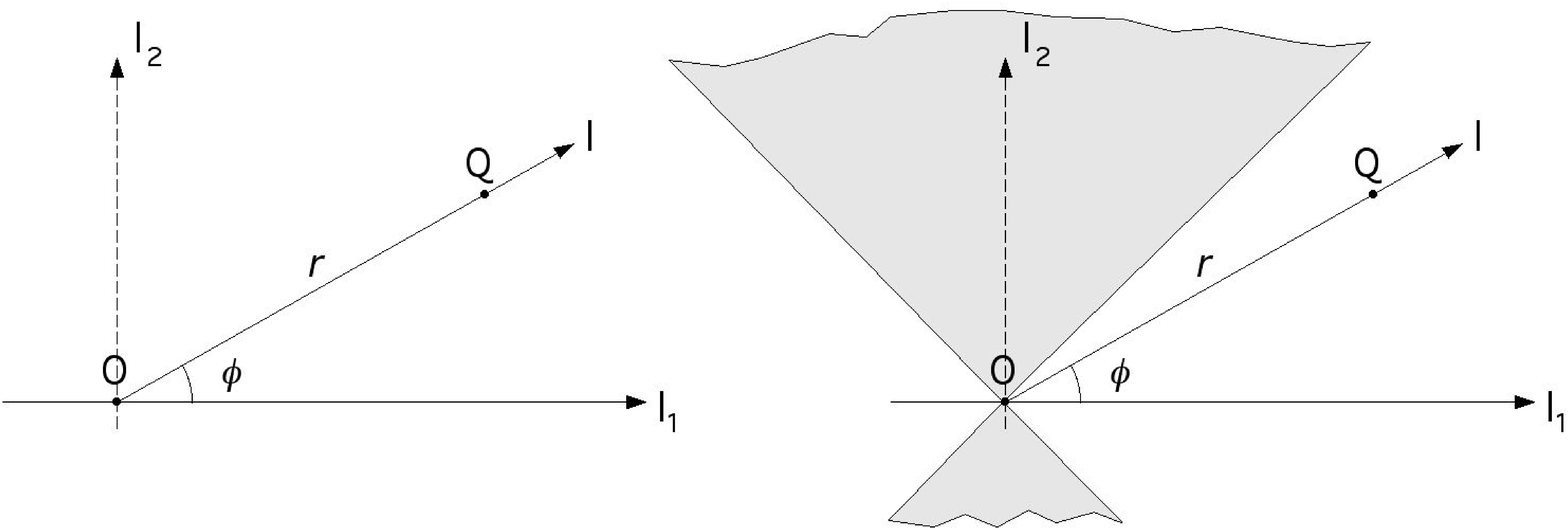}\hfill}

\noindent{\sc Figure 5 ab}.{\enskip} The `polar' coordinates $(r,
\phi)$. The diagram depicts the geometrical meaning of polar
coordinates $(r, \phi)$ in a general CK space $\CKspace$, both in
the locally Riemannian case $\kii\>0$  (left) and in the
pseudo-Riemannian case $\kii\<0$ (right). In all cases $l_1, l_2,
l$ are geodesics, and $l_1, l_2$ are orthogonal. The light cone
through $O$ is also shown in the lorentzian diagram. The
coordinate  $r$ has label $\ki$ while $\phi$ has label $\kii$. In
the Riemannian case, the coordinate $r$ is non-negative, only
vanishes at point $O$, where polar coordinates are singular, and
the angular coordinate $\phi$ ranges in the interval $[0,
2\pi/\sqrt{\kii}]$ with the usual periodic conditions. In the
pseudo-Riemannian case $r$ vanishes along the isotropes through
$O$ (and would be pure imaginary in the shaded area with
space-like separation to $O$); the angle $\phi$ ranges in the
interval $[-\infty, \infty]$ and for a given $\phi$ the natural
range of $r$ involves positive as well as negative values.

\bigskip\bigskip\bigskip

{\hfill\includegraphics[width=400pt]{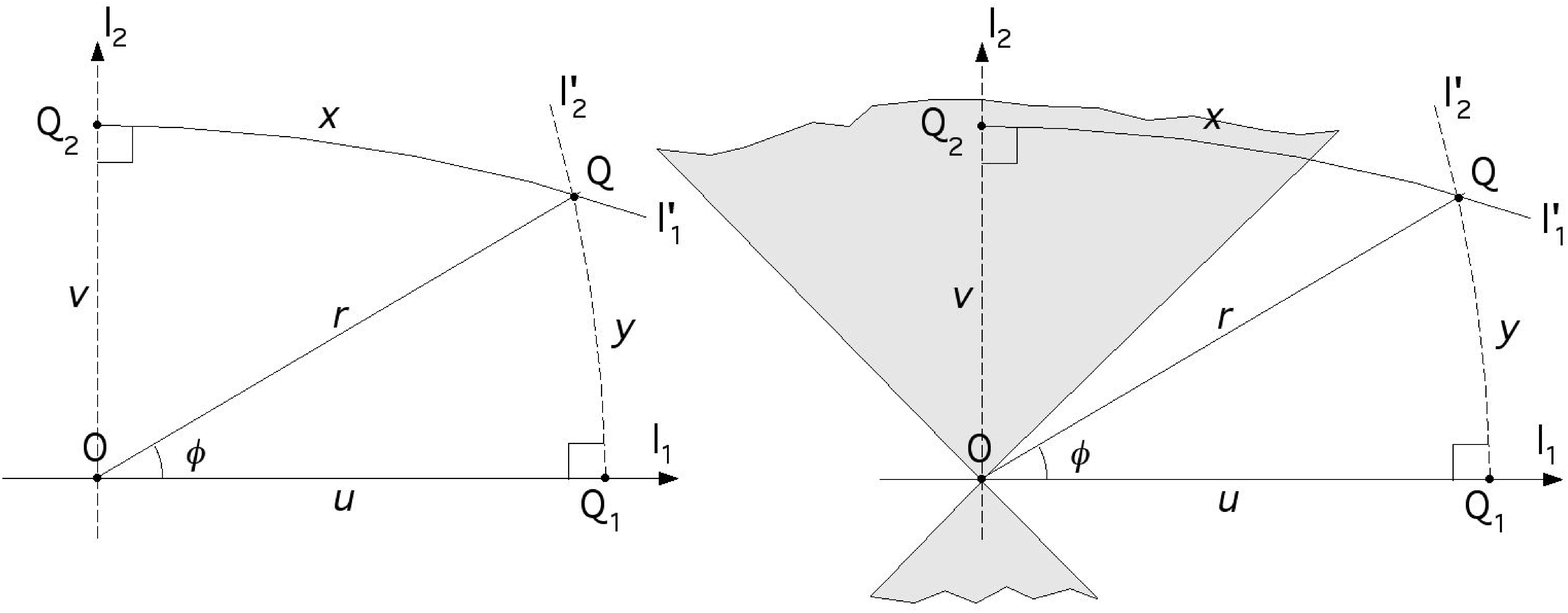}\hfill}

\noindent{\sc Figure 6 ab}.{\enskip} The `parallel' coordinates
$(u, y)$ and $(v,x)$. The diagram depicts the geometrical meaning
of the coordinates $(u, y)$ and $(v,x)$, for the same situation
and with the same conventions as in Fig.5. The lines $l_1', l_2'$
are geodesics through $Q$ orthogonal to $l_2, l_1$ respectively.
The coordinates $u, x$ have label $\ki$ and are (locally) defined
near $O$ in both the Riemannian and pseudo-Riemannian cases. The
coordinated $v, y$ have label $\ki\kii$ and the corresponding
geodesics are represented dashed; in the pseudo-Riemannian case
this means these geodesics are space-like. In all cases the
ordinary sign convention applies. When $\ki\neq0$, $x\neq u$ and
$v\neq y$, and equality is a degenerate property of the flat case.
See the text in the appendix for more details, and note that the
natural interpretation of all coordinates is as canonical
parameters of one-parameter subgroup of translations along the
lines $l_1, l_2, l_1', l_2'$ or of rotations around the point $O$.


\end{document}